\documentclass[aps,prx,twocolumn,showpacs,superscriptaddress,groupedaddress]{revtex4}  
\usepackage{graphicx}  
\usepackage{dcolumn}   
\usepackage{bm}        
\usepackage{amssymb}   
\usepackage{mathrsfs}
\usepackage{xcolor}

\definecolor{tab_blue}{HTML}{1F77B4}
\definecolor{tab_orange}{HTML}{FF7F0E}
\definecolor{tab_green}{HTML}{2CA02C}
\definecolor{tab_red}{HTML}{D62728}
\definecolor{tab_purple}{HTML}{9467BD}
\definecolor{tab_brown}{HTML}{8C564B}
\definecolor{tab_pink}{HTML}{E377C2}
\definecolor{tab_gray}{HTML}{7F7F7F}
\definecolor{tab_olive}{HTML}{BCBD22}
\definecolor{tab_cyan}{HTML}{17BECF}

\def\avg#1{\big <#1 \big >}
\def\lesssim{\ \raise.3ex\hbox{$<$}\kern-0.8em\lower.7ex\hbox{$\sim$}\ }
\def\gesim{\ \raise.3ex\hbox{$>$}\kern-0.8em\lower.7ex\hbox{$\sim$}\ }

\begin{document}
\widetext
\title{Critical fluctuations at a many-body exceptional point}
\author{Ryo Hanai}
\email{rhanai@uchicago.edu}
\affiliation{James Franck Institute and Department of Physics, University of Chicago, Illinois, 60637, USA} 
\affiliation{Department of Physics, Osaka University, Toyonaka 560-0043, Japan} 
\author{Peter B. Littlewood}
\affiliation{James Franck Institute and Department of Physics, University of Chicago, Illinois, 60637, USA} 
\affiliation{Materials Science Division, Argonne National Laboratory, Argonne, Illinois 60439, USA} 
\date{\today}
\begin{abstract}
Critical phenomena arise ubiquitously in various context of physics, from condensed matter, high energy physics, cosmology, to biological systems, and consist of slow and long-distance fluctuations near a phase transition or critical point. Usually, these phenomena are associated with the softening of a massive mode. Here we show that a novel, non-Hermitian-induced mechanism of critical phenomena that do not fall into this class can arise in the steady state of generic driven-dissipative many-body systems with coupled binary order parameters such as exciton-polariton condensates and driven-dissipative Bose-Einstein condensates in a double-well potential. The criticality of this ``critical exceptional point'' is attributed to the \textit{coalescence} of the collective eigenmodes that convert all the thermal-and-dissipative-noise activated fluctuations to the Goldstone mode, leading to anomalously giant phase fluctuations that diverge at spatial dimensions $d\le 4$. 
Our dynamic renormalization group analysis shows that this gives rise to a strong-coupling fixed point at dimensions as high as $d<8$ associated with a new universality class beyond the classification by Hohenberg and Halperin, indicating how anomalously strong the many-body corrections are at this point.
We find that this anomalous enhancement of many-body correlation is due to the appearance of a sound mode at the critical exceptional point despite the system's dissipative character.
\end{abstract}
\maketitle

\section{Introduction}

Understanding and manipulating dissipation effects in open quantum systems \cite{Weiss} is increasing in importance, due to its crucial role in designing new optical devices and performing quantum computation.  
Particularly intense interest has recently emerged to the study of `exceptional points (EPs)' \cite{Kato1966,Bender1998,Heiss1999} that can arise in these dissipative devices.
An EP is a point where two (or more) eigenstates that characterize the dynamics coalesce owing to the non-Hermitian nature of the system, such that they lose their completeness, leading to a spectral singularity. 
It turns out that this singularity gives rise to a number of counter-intuitive phenomena in the vicinity of the EP, such as loss-induced transmission \cite{Guo2009}, unidirectional invisibility \cite{Lin2011}, enhanced quantum sensitivity \cite{Chan2017,Hodaei2017} and chiral behavior \cite{Doppler2017}. 
These concepts have been proven to be applicable to a rich variety of many-body systems such as superconductivity \cite{Chtchelkatchev2012}, atomic gases \cite{Li2019}, spin-chains \cite{Luitz2019}, and correlated materials \cite{Kozii2017}, where EPs arise in the spectral properties of these systems. 


\begin{figure*}
\begin{center}
\includegraphics[width=0.6\linewidth,keepaspectratio]{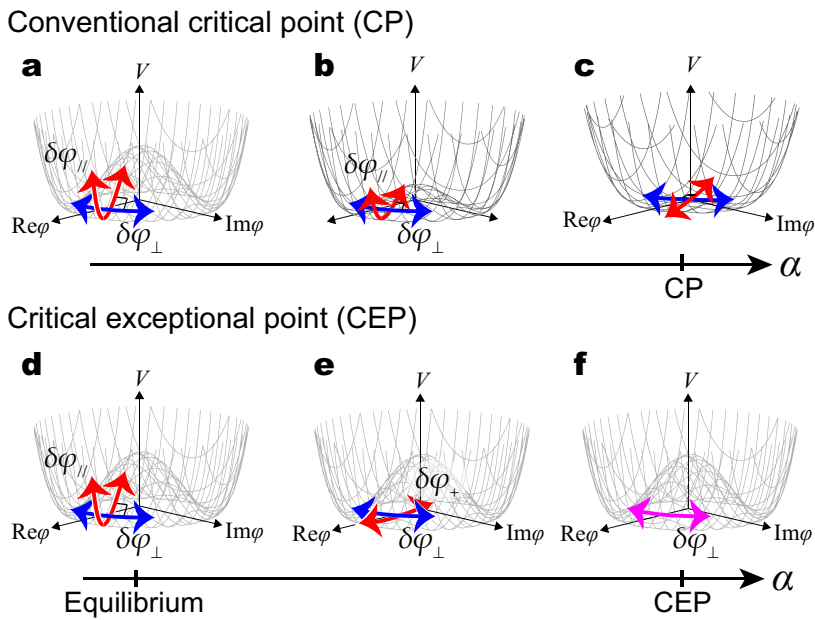}
\end{center}
\caption{
(Color online) 
Difference between the conventional critical point (CP) and the critical exceptional point (CEP).
(a)-(c) Schematic explanation of the criticality of the CP in conventional equilibrium cases (the figure describes the case of $U(1)$-symmetry breaking transition). As one sweeps the parameter $\alpha$ to the CP, the massive longitudinal mode $\delta\varphi_{\parallel}$ itself softens at the CP caused by the flattening of the free energy landscape $V$ (where $\varphi$ is the order parameter). Note how the two eigenmodes, the Goldstone mode $\delta\varphi_{\perp}$ and the longitudinal mode $\delta\varphi_{\parallel}$, are always orthogonal. 
(d)-(f) Schematic explanation of the criticality of the CEP. In the non-Hermitian case, the two eigenmodes are not necessarily orthogonal. As a result, while the Goldstone's theorem ensures the Goldstone mode $\delta\varphi_\perp$ to be an eigenmode, the other eigenmode $\delta\varphi_+$ is \textit{not} pointing to the longitudinal direction. At the CEP, where the two eigenmodes coalesce to the Goldstone mode, anomolously giant phase fluctuations occur which diverges at $d\le 4$, as well as anomalously enhanced many-body correlation effect.  
(We briefly note that the free energy drawn in Figs. \ref{fig_critical}(d)-(f) is schematic.)
}
\label{fig_critical}
\end{figure*}

One of the most striking phenomena that arise in many-body systems are critical phenomena, 
which are collective many-body phenomena associated with the divergence of length and time scales near a phase transition or critical point  \cite{Amit2005}.
Critical points (CPs) and EPs have a crucial feature in common: the occurrence of a gap closure \cite{Altland_Simons}, 
where in the latter the energy difference between the two eigenstates vanishes as they coalesce.
This raises the issue whether EPs can possess similar properties to CPs such as critical fluctuations when EPs occur in a many-body context. 
Indeed, some EPs of a non-Hermitian many-body Hamiltonian are shown to be marked as a quantum critical point \cite{Tripathi2016,Poccia2015,Ashida2017}.
However, these were found only in the lowest-energy eigenstate, limiting to situations where the thermal or dissipation-induced noise is negligible or can be excluded by post-selection.

Here we propose a novel, non-Hermitian induced critical phenomena \textit{activated} by the thermal and dissipative noise that occurs at a many-body EP, which can arise in generic driven-dissipative quantum many-body systems composed of coupled binary order parameters.
We illustrate this phenomenon by analyzing a two-component driven-dissipative 
condensate, where examples in scope include exciton-polariton condensates \cite{Carusotto2013} (composed of excitons and photons) and driven-dissipative Bose-Einstein condensates (BECs) in a double-well potential \cite{Graefe2012,Cartarius2012,Dast2013}. 
Strinkingly, our analysis reveal that our many-body EP (which we refer to as critical EP, CEP) exhibits anomalously giant phase fluctuations that diverge at spatial dimensions $d\le 4$, which is to be compared to the conventional case where the divergence only happens at $d\le 2$ \cite{Mermin1966,Hohenberg1967}).
We also find the emergence of a \textit{sound} mode despite the presence of the dissipation, which turns out to anomalously enhance many-body correlation effects that become relevant at dimensions as high as $d<8$ (which is to be compared to, e.g., the Ising model or XY model where the many-body correlation effect is relevant at $d< 4$).
We demonstrate this physics by performing a dynamic renormalization group analysis \cite{Hohenberg1977} to identify a strong-coupling fixed point in the vicinity of $d_c=8$  dimensions associated with a new universality class beyond the classification by Hohenberg and Halperin \cite{Hohenberg1977}.


These anomalous critical fluctuations are caused by a fundamentally different mechanism of criticality from the conventional critical phenomena. 
At conventional CPs, the critical fluctuations arise due to the softening of the longitudinal mode, originated from the flattening of the free energy landscape (Figs. \ref{fig_critical}(a)-(c)); longitudinal and transverse fluctuations are separable.
In the driven-dissipative case, on the other hand, these collective eigenmodes need not be orthogonal, because of the non-Hermitian nature of the system (Figs. \ref{fig_critical}(d),(e)).
We show here that these collective eigenmodes \textit{coalesce} (i.e. they become collinear) at the CEP  (Fig. \ref{fig_critical}(f)); hence thermally activated fluctuations are all converted to the Goldstone mode. 
This peculiar property turns out to give rise to the anomalously giant phase fluctuations, which is reminiscent of the noise-induced pattern formation due to the non-orthogonality of eigenstates discussed in Ref. \cite{Biancalani2017}.
We also find that this mode coalescence gives rise to a sound mode -- despite the dynamics are overdamped by the dissipation and thus usually gives a diffusive mode \cite{Wouters2007,Szymanska2006} -- that turns out to lead to the anomalous enhancement of many-body correlation effects. 


It is worth noting that our CEP is found in the \textit{steady state} that characterizes the longtime behavior of the driven-dissipative many-body system, in stark contrast to Refs. \cite{Tripathi2016,Poccia2015,Ashida2017} where the lowest-energy state (that exhibits criticality) needs to be postselected after measurement since the system would eventually approach a trivial state such as infinite temperature or zero-particle state by the heating or particle loss processes.
We also remark that, since the critical fluctuations of our system are activated by the thermal and dissipative noise, our CEP can be regarded as a \textit{semiclassical} and \textit{dynamical} critical phenomena, which is a different class of criticality from the quantum and static critical phenomena at many-body EPs found in the previous studies \cite{Tripathi2016,Poccia2015,Ashida2017}.


The discovered mechanism is generic;  
the CEP should arise as long as the many-body system is (1) driven-dissipative, (2) composed of two components, and (3) exhibit spontaneous symmetry breaking, since the rise of our CEP is ultimately due to the non-orthogonality of collective eigenmodes. 
Our previous work \cite{Hanai2019} has shown that driven-dissipative systems composed of two components exhibit a non-Hermitian phase transition with an endpoint of its phase boundary marked by a many-body EP (as schematically shown and demonstrated in Fig. \ref{fig_phasediagram}), proposed as a new interpretation of the phase transition observed in some polariton experiments  in the $U(1)$-broken phase \cite{Bajoni2008,Balili2009,Nelsen2009,Tempel2012a,Tsotsis2012,Horikiri2013})
 (the so-called ``second threshold''). 
The two-component nature offers two branches of eigenstates that the system can condense into, letting us define two distinct phases of matter not distinguished by symmetry -- analogous to liquid and gas in an equilibrium system. 
Because of the driven-dissipative feature, the CEP arises as a point where the two phases coalesce, making that the endpoint of the phase boundary \cite{Hanai2019}.
(See Appendix \ref{appendix NESS} for details.)

Due to the generality of its mechanism, it should be possible to implement the CEP in generic driven-dissipative systems by engineering the system to be binary, e.g. by preparing a double-well potential or checkerboard lattice structure of gain and loss components.
Considering the rapid experimental development in various driven-dissipative many-body platforms, such as ultracold atoms \cite{Daley2014}, circuit QED \cite{Houck2012,Fitzpatrick2017}, photon BEC \cite{Klaers2010}, plasmonic-lattice polariton BEC \cite{Hakala2018}, and strongly interacting photons \cite{Ma2019}, it seems promising to realize the CEP in these systems. 
An exciton-polariton condensate \cite{Carusotto2013} and a plasmonic-lattice polariton BEC \cite{Hakala2018} 
seems especially promising, since they are both composed of two components (i.e., photons and excitons for the former and surface lattice resonance mode and the dye molecule excitation modes for the latter) and thus meet all three criteria raised above. 
Especially in the polariton system, both the first-order-like phase transition \cite{Bajoni2008,Balili2009,Nelsen2009,Tempel2012a,Tsotsis2012,Horikiri2013})
(corresponding to the solid line in Fig. \ref{fig_phasediagram}(a)) and the crossover behavior \cite{Deng2003} has already been observed \cite{Hanai2019}. 
This makes us strongly expect that the realization of the CEP can be achieved with the current experimental techniques, by tuning the pump power and detuning.
(See the discussion in the Supplemental Material in Ref. \cite{Hanai2019} for further details on comparison to polariton experiments.)

The appearance of the CEP does not rely on the fact the system is composed of quantum particles; analogous phenomena may appear even in \textit{classical} many-body systems as well, when the system is intrinsically driven out of equilibrium, as in active matter systems \cite{Marchetti2013}.  
Indeed, recently, we have found the emergence an analogous point  \cite{Fruchart2020} in a generalized Vicsek model of flocking \cite{Vicsek1995,Toner1995} and Kuramoto model of synchronization \cite{Kuramoto1984} with non-reciprocally interacting (classical) agents. There, the CEP was found to arise as a phase transition point to a time crystal phase. 
A similar point has been found in the neural nets of synaptically coupled excitatory and inhibitory neurons in the neocortex \cite{Cowan2016} (where the CEP is phrased as the Bogdanov-Takens point).
We expect to find more in the near future. 

\section{Noisy driven-dissipative Gross-Pitaevskii equation for binary condensates}

\begin{figure}
\begin{center}
\includegraphics[width=0.85\linewidth,keepaspectratio]{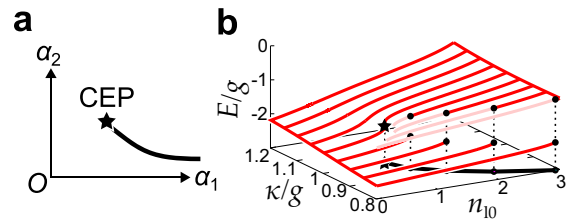}
\end{center}
\caption{
(Color online) 
Non-Hermitian phase transition and the critical exceptional point (CEP).
(a) Schematic phase diagram of driven-dissipative many-body systems with coupled binary order parameters, in terms of the generic input parameters $(\alpha_1,\alpha_2)$. The solid line is the phase boundary of the non-Hermitian induced phase transition, which exhibits an endpoint at the critical exceptional point (CEP). 
(b) Demonstration of the appearance of the CEP for systems described by the coupled driven-dissipative Gross-Pitaevskii equations. At a small decay rate $\kappa/g<1$, as the density of the loss component $n_{\rm l0}=|\Phi_{\rm l}^0|^2$ increases, the steady state exhibit a phase transition signaled by a discontinuity in the condensate emission energy $E$. The CEP found at $\kappa/g=1$, represented by the star, marks the endpoint of the phase boundary. The discontinuity is absent at $\kappa/g>1$. We set $\omega_{\rm l}/g=-1.9, \omega_{\rm g}/g=-2$, and $U_{\rm g}/g=0.1$. 
}
\label{fig_phasediagram}
\end{figure}

Below, we consider a driven-dissipative, repulsively interacting BEC composed of two components. 
Examples of such systems include exciton-polariton condensates, driven-dissipative condensates in a double-well potential, plasmonic-lattice polariton condensates, etc.
To our knowledge, this is the simplest system that exhibits a non-Hermitian induced phase transition associated with the rise of the CEP \cite{Hanai2019}.


Our goal is to reveal the dynamic critical properties of the CEP. 
For a one-component case, it has been shown \cite{Sieberer2016} by coarse-graining the Keldysh partition function \cite{Altland_Simons} that the critical properties of a driven-dissipative condensate can be captured by taking the noise average of the noisy driven-dissipative Gross-Pitaevskii (GP) equation \cite{Wouters2007,Szymanska2006}.
It is straightforward to extend this discussion to our two component case \cite{Hanai2019},  to show that the stochastic equations of motion to consider is  
\begin{eqnarray}
i\partial_t \bm \Psi(\bm r,t) 
=A_{\rm GP}(\nabla^2)\bm \Psi(\bm r,t)+\bm\eta(\bm r,t),
\label{GP}
\end{eqnarray}
with 
$\bm \Psi=(\Psi_{\rm l},\Psi_{\rm g})^{\mathsf T}$, 
$\bm \eta=(\eta_{\rm l},\eta_{\rm g})^{\mathsf T}$,
and 
\begin{eqnarray*}
&&A_{\rm GP}(\nabla^2)
\nonumber\\
&&=\left(\begin{array}{cc}
\omega_{\rm l}-i\kappa- K_{\rm l}\nabla^2
& g \\
g & \omega_{\rm g} +iP -\tilde K_{\rm g}\nabla^2+\tilde U_{\rm g}|\Psi_{\rm g}|^2
\end{array}\right),
\end{eqnarray*}
where $\alpha={\rm l(g)}$ labels the loss (gain) component with its net one-body decay (gain) rate given by $\kappa (P)$. 
Here, $\omega_{\rm l(g)}$ and $\Psi_{\rm l(g)}(\bm r,t)$ are the energy and macroscopic wave function of the loss (gain) component, respectively. 
$g$ is the inter-component coupling and $\tilde K_{\rm g}=K_{\rm g}-iD_{\rm g}, \tilde U_{\rm g}=U_{\rm g}-iv_{\rm g}$  are complex coefficients, where $D_{\rm g},U_{\rm g},v_{\rm g}$ are the diffusion constant, repulsive interaction strength, and two-body loss rate that gives nonlinear saturation of the gain component, respectively, 
and $K_{\rm l(g)}$ determines the kinetics of the loss (gain) component.
We have assumed in this model that only the gain components interact and have a nonlinear saturation, which however, does not affect the critical properties of the CEP. 
White noise $\eta_\alpha(\bm r,t)$ that originates from thermal fluctuations and dissipation is characterized by
$\avg{\eta_\alpha^*(\bm r,t)\eta_\beta(\bm r',t')}
=\Delta_{\alpha}\delta_{\alpha\beta}\delta^d(\bm r-\bm r')\delta(t-t')$, and
$\avg{\eta_\alpha(\bm r,t)}=\avg{\eta_\alpha(\bm r,t)\eta_\beta(\bm r',t')}=0$, 
where the noise level $\Delta_{\alpha}$ is determined from the sum of the one-body gain and loss rates of respective components \cite{Sieberer2016}. 
In general, the noise may not satisfy the dissipation-fluctuation theorem in driven-dissipative systems.

We stress that we are interested in the \textit{collective} excitation properties of the \textit{steady state} (enabled to stabilize thanks to the nonlinear saturation effect), in stark contrast to the majority of problems discussed in the field of non-Hermitian physics where transient dynamics of the excited states are often discussed \cite{Konotop2016}.
Despite its  similarity, the consequences are crucially different \cite{Hanai2019} where it exhibits a first-order-like phase transition associated with a critical point as shown and demonstrated in Fig. \ref{fig_phasediagram}. 
(See also the discussion below and Appendix \ref{appendix NESS}.)
We will see in the following that anomalous critical phenomena arises at this point, which is absent in the conventional EPs. 
We also emphasize that this stochastic equation of motion incorporates beyond-meanfield effects, due to the white noise that continuously fluctuate the condensate about the steady state.

Before going into the analysis of critical fluctuations, we first give a brief review on the steady state properties of this system in the absence of noise $\eta_\alpha(\bm r,t)$ \cite{Hanai2019}.
At low pump power $P(<P_{th})$ where the gain is not large enough to compensate the loss, the macroscopic wave function dies out to a normal state, i.e., $\Psi_\alpha(t\rightarrow\infty)=0$.
On the other hand, 
at a sufficiently high pump power $P>P_{th}$, 
the amplitude of the macroscopic wave function starts to 
grow until the nonlinear saturation makes the system  
approach a uniform steady state.
The macroscopic wave function $\Psi_\alpha^0(t)
\equiv\Phi^0_\alpha e^{-iEt}$ then follows the relation
\begin{eqnarray}
E \Phi^0_\alpha 
= \sum_{\beta={\rm l,g}}[A_{\rm GP}(\Phi^0_{\rm l},\Phi^0_{\rm g})]_{\alpha\beta}\Phi^0_\beta,
\end{eqnarray}
where $E$ is the (real) condensate emission energy.
This steady state condition determines the amplitude of $\Phi^0_\alpha$ for a given setup. 
We emphasize that the above equation is essentially different from conventional non-Hermitian quantum mechanics problems \cite{Konotop2016}, where the nonlinear effect (that makes it possible to reach a steady state) is present and the emission energy $E$, corresponding to the eigenvalue, is necessarily real because the system is in the steady state. 
As detailed in Appendix \ref{appendix NESS}, the above steady-state restriction allows us to classify the solutions into two types \cite{Hanai2019}, that satisfies $E=E_-$ and $E=E_+$, respectively, where $E_\pm$ are the two eigenvalues of the matrix $A_{GP}$. These physically correspond to condensation into the lower and upper branch, respectively.

Crucially, because of the non-Hermitian nature of $A_{\rm GP}$, the two eigenvectors of $A_{\rm GP}$ need not be orthogonal, giving rise to a point where the two eigenvectors  coalesce; the CEP. 
This corresponds to the coalescence of the two solutions. 
The CEP is proven to mark the endpoint of the first-order-like phase boundary that can occur  between the two types of solutions \cite{Hanai2019}, in a similar manner to the critical point of a liquid and gas where the two phases also coalesce. 
This is demonstrated in Fig. \ref{fig_phasediagram}(b).
The CEP is found at the parameter that satisfies $\kappa=g$  and $\omega_{\rm l}=\omega_{\rm g}+U_{\rm g}|\Phi_{\rm g}^0|^2$. (See the Appendix \ref{appendix NESS} for details.) 
We note that there are only two parameters to tune, which is the same number of parameters for the conventional EP and critical point of liquid-gas phase diagram to be realized.

For the investigation of criticality, it is useful to rewrite the GP equation (\ref{GP}) in terms of the noise-activated amplitude $\delta|\Phi_\alpha(\bm r,t)|$ and phase fluctuations $\delta\theta_\alpha(\bm r,t)$ by rewriting the macroscopic wavefunction as the deviation from the steady state, $\Psi_\alpha(\bm r,t)=[\Phi^0_\alpha+\delta\Phi_\alpha(\bm r,t)]e^{-iEt}
=(|\Phi_\alpha^0|+\delta|\Phi_\alpha(\bm r,t)|)e^{i(\theta^0_\alpha+\delta\theta_\alpha(\bm r,t))} e^{-iEt}$, where $\Phi_\alpha^0=|\Phi_\alpha^0| e^{i\theta^0_\alpha}$.
By integrating out $\delta|\Phi_\alpha(\bm r,t)|$ up to linear order and further assuming that the amplitude dynamics are overdamped \cite{Altman2015}, we arrive at a Kardar-Parisi-Zhang (KPZ)-like \cite{Kardar1986} stochastic equation of motion (See Appendix \ref{appendixKPZ}.),  
\begin{eqnarray}
&&\partial_t
\delta\theta_\alpha(\bm r,t) 
=\sum_{\beta={\rm l,g}}W_{\alpha\beta}(\nabla)
\delta\theta_\beta(\bm r,t)
+t_\alpha(\Delta\delta\theta(\bm r,t))^2
\nonumber\\
&& 
+\sum_{\beta,\gamma={\rm l,g}}\lambda^\alpha_{\beta\gamma}
(\nabla\delta\theta_\beta(\bm r,t))
\cdot
(\nabla\delta\theta_\gamma(\bm r,t))+
\xi_\alpha (\bm r,t) ,
\label{coupledKPZ}
\end{eqnarray}
with
\begin{eqnarray*}
W(\nabla)
=
\left(\begin{array}{cc}
-s_{\rm l}+\nu_{\rm ll}\nabla^2 
& s_{\rm l}+\nu_{\rm lg}\nabla^2 \\
-s_{\rm g}+\nu_{\rm gl}\nabla^2 
& s_{\rm g}+\nu_{\rm gg}\nabla^2
\end{array}\right),
\end{eqnarray*}
and $\Delta\delta\theta(\bm r,t)=\delta\theta_{\rm l}(\bm r,t)-\delta\theta_{\rm g}(\bm r,t)$. 
Here, we have retained the most relevant nonlinear terms and 
 the real parameters $s_\alpha, \nu_{\alpha\beta},\lambda^\alpha_{\beta\gamma}$ are determined from the parameters used in $A_{\rm GP}$, 
where its explicit form is given in Appendix \ref{appendixKPZ}.
$\xi_\alpha(\bm r,t)$ is a white noise for the phases, characterized by $\avg{\xi_\alpha(\bm r,t)\xi_\beta(\bm r',t')}
=\sigma_{\alpha\beta}\delta^d(\bm r-\bm r')\delta(t-t')$
and $\avg{\xi_\alpha(\bm r,t)}=0$.

\section{Anomalously giant phase fluctuations at the CEP}
\label{sec linear}

We show below that an anomalous critical behavior appears at the CEP. At the CEP, we find $s_{\rm l}=s_{\rm g}=\kappa$, as shown in Appendix \ref{appendixKPZ}.  
Let us start with a linearized theory (i.e. $t_\alpha=\lambda_{\alpha\beta}=0$). 
By solving the secular equation ${\rm det}[-i\omega\bm 1-W(\bm k)]=0$, the eigenenergies are given by
\begin{eqnarray}
\omega_\pm (\bm k)=
\frac{1}{2}\big[
-i(\gamma + 2D\bm k^2)\pm \sqrt{-\gamma^2+4v^2 \bm k^2}
\big],
\label{eigenenergy}
\end{eqnarray}
where $\gamma=s_{\rm l}-s_{\rm g},v^2=[2(s_{\rm g}\nu_{\rm lg}+s_{\rm l}\nu_{\rm gl})+(s_{\rm l}+s_{\rm g})(\nu_{\rm gg}-\nu_{\rm ll})]/2$ and $D=(\nu_{\rm gg}+\nu_{\rm ll})/2$. 
We remark that $\gamma$ and $v^2$ can take a negative value, which may cause a dynamical instability, not uncommon for systems with loss and gain.
In this paper, we assume that those situations are avoided by having a large enough nonlinear saturation $v_{\rm g}$ and diffusion constant $D_{\rm g}$ in the gain component, to have $\gamma,v^2\ge0$. 
(See the discussions in Appendix \ref{appendixKPZ}.)

When the system is away from the CEP, i.e. $\gamma > 0$, the eigenmodes are given by the diffusive Goldstone mode $\omega_-(\bm k) \propto -i \bm k^2$ \cite{Szymanska2006,Wouters2007} and a relaxational mode $\omega_+(\bm k)=-i \gamma$. 
Since the relaxational mode would be gapped away as we further coarse-grain the system and thus play no role in the effective low-energy physics, 
the physics is essentially the same as the one-component case and recovers the dynamic scaling behavior  \cite{Altman2015,Sieberer2016,Wachtel2017,He2017,Comaron2018}  known to obey the KPZ scaling \cite{Kardar1986}.  

The situation is dramatically different at the CEP
$\gamma=0$. In this case, we find both eigenmodes $\omega_\pm(\bm k)$ to be gapless, which interestingly are \textit{sound} modes, 
\begin{eqnarray}
\omega_\pm (\bm k)=\pm v|\bm k|-i D\bm k^2,
\label{GoldstoneCEP}
\end{eqnarray}
showing that both components play role and thus modifies the scaling properties.

A crucial observation is that, this gap closure is associated with the \textit{coalescence} of the collective modes, which is crucially different from just being degenerate.
The eigenmodes in the uniform limit $\bm k\rightarrow 0$  are given by, 
\begin{eqnarray}
\left(\begin{array}{c}
\delta\theta_{\rm l} \\
\delta\theta_{\rm g}
\end{array}\right )
\propto
\left(\begin{array}{c}
1 \\ 
1
\end{array}\right ), 
\left(\begin{array}{c}
\delta\theta_{\rm l} \\
\delta\theta_{\rm g}
\end{array}\right )
\propto
\left(\begin{array}{c}
s_{\rm l}/s_{\rm g} \\ 
1
\end{array}\right ),
\label{eigenmodes}
\end{eqnarray}
for the corresponding eigenenergies $\omega_-(\bm k)$ and $\omega_+(\bm k)$, respectively. The former  in-phase mode, i.e. the Goldstone mode, is assured to be gapless by the global symmetry under the transformation $\Psi_\alpha(\bm r,t)\rightarrow\Psi_\alpha(\bm r,t)e^{i\phi}$.
These modes coalesce at the CEP $s_{\rm l}=s_{\rm g}$ as schematically described in Figs. \ref{fig_critical}(d)-(f), giving rise to the gap closure. 
This gap-closing mechanism is fundamentally different from that in the conventional CPs, where the longitudinal mode itself softens by the flattening of the free energy but is still orthogonal to the Goldstone mode (Figs. \ref{fig_critical}(a)-(c)).

The rise of the sound mode at the CEP (Eq. (\ref{GoldstoneCEP})) is one of the outcome of this mode coalescence. 
Noting that $\hat W(\bm k=0)$ is at an exceptional point at the CEP, the linear $|\bm k|$ dependence in the dispersion originates from the square-root behavior in Eq. (\ref{GoldstoneCEP}), a typical property seen in the vicinity of the exceptional point \cite{Chan2017,Hodaei2017}. 
In fact, the appearance of the sound mode plays a crucial role in exhibiting anomalously large many-body correlation effects, as we show in the next section.

This peculiar collective mode-coalescence  at the CEP gives rise to anomalously giant phase fluctuations. 
To see this in a transparent way, it is useful to \textit{triangularize} the kernel $W(\bm k)$ by using an \textit{orthogonal} basis $U^{\dagger}(\bm k)\equiv(\bm u_\perp(\bm k),\bm u_\parallel(\bm k))^{\mathsf T}$
(instead of diagonalizing $W(\bm k)$, which is ill-defined at $\bm k\rightarrow 0$ at the CEP)  as,
\begin{eqnarray*}
\bar W(\bm k)
= U(\bm k)W(\bm k)U^{\dagger}(\bm k)=
\left(\begin{array}{cc}
-i\omega_-(\bm k) & \zeta \\
0 &-i \omega_+(\bm k)
\end{array}\right).
\end{eqnarray*}
Here, we have chosen one basis vector as the Goldstone mode, $W(\bm k)\bm u_\perp(\bm k)=-i\omega_-(\bm k)\bm u_\perp(\bm k)$, and the other basis, satisfying $W(\bm k)\bm u_\parallel(\bm k)=-i\omega_+(\bm k)\bm u_\parallel(\bm k)+\zeta\bm u_\perp(\bm k)$, is chosen as the longitudinal direction that is perpendicular to the Goldstone mode $\bm u_\parallel(\bm k)\cdot\bm u_\perp(\bm k) = 0$. 
Importantly, the off-diagonal piece $\zeta=s_{\rm l}+s_{\rm g}$ arises from the non-Hermitian nature of $W$, which converts the longitudinal fluctuations to the Goldstone mode.

The Green's function  in this basis 
$\bar G_{ss'}^0(\bm k,\omega)=([-i\omega \bm 1-\bar W(\bm k)]^{-1})_{ss'}$ is given by, 
\begin{eqnarray}
\bar G^0(\bm k,\omega)=
\left(
\begin{array}{cc}
\frac{i}{\omega-\omega_-(\bm k)} & 
\frac{-\zeta}{(\omega-\omega_-(\bm k))(\omega-\omega_+(\bm k))}\\
0 & \frac{i}{\omega-\omega_+(\bm k)} 
\end{array}
\right).
\label{Green}
\end{eqnarray}
Here, we find that the off-diagonal, non-Hermitian induced component $\bar G_{\perp\parallel}^0$ exhibits a peculiarly strong singularity at the CEP, involving two gapless poles at $\omega=\omega_\pm(\bm k)$. 
As a result, noting that the phase fluctuations are expressed in terms of Green's functions as $\delta\bar\theta_{s=\perp,\parallel}(\bm k,\omega)=\sum_{s'=\perp,\parallel}\bar G_{ss'}^0(\bm k,\omega)\bar\xi_{s'}(\bm k,\omega)$
(where $\delta\bar \theta_{s}(\bm k,\omega) = \sum_{\alpha={\rm g,l}}U_{s\alpha}(\bm k)\delta\theta_\alpha(\bm k,\omega)$ and 
$\bar\xi_{s}(\bm k,\omega) = \sum_{\alpha={\rm g,l}}U_{s\alpha}(\bm k)\xi_\alpha(\bm k,\omega)$),
the most strongly diverging term in the equal-time correlation function involves two $\bar G_{\perp\parallel}^0$'s, 
\begin{eqnarray}
&&\avg{\delta\theta_{\alpha(={\rm l,g})}(\bm r)
\delta\theta_{\beta(={\rm l,g})}(\bm r'))}
\sim
\int_0^{\Lambda_c} d k k^{d-1}
e^{i\bm k\cdot (\bm r-\bm r')}
\nonumber\\
&&\times \int_{-\infty}^\infty \frac{d\omega}{2\pi}
\bar G_{\perp\parallel}^0(\bm k,\omega)
\sigma_{\parallel\parallel}
\bar G_{\perp\parallel}^0(-\bm k,-\omega)
\nonumber\\
&&\sim 
\int_0^{\Lambda_c} d k k^{d-1}
e^{i\bm k\cdot (\bm r-\bm r')}
\frac{A}{k^4},
\label{fluctuation_CEP}
\end{eqnarray}
which diverges for $d\le 4$,
implying the absence of long-range order at the CEP.
Here,  `$\sim$' means that we have retained the most strongly diverging term and 
\begin{eqnarray}
A\equiv\frac{\kappa^2\sigma_{\parallel\parallel}}{Dv^2},
\label{def A}
\end{eqnarray}
with longitudinal noise $\sigma_{\parallel\parallel}=(\sigma_{\rm ll}+\sigma_{\rm gg}-\sigma_{\rm lg}-\sigma_{\rm gl})/2$. 
(We have used the relation $\zeta=2\kappa$ that holds at the CEP.)
$\Lambda_c$ is an ultraviolet cutoff.
The obtained phase fluctuations are anomalously giant, in the sense that the phase fluctuations in the conventional case 
$\avg{(\delta\theta)^2}\sim \int_0^{\Lambda_c} dk k^{d-1}\cdot k^{-2}$ diverges only for $d\le 2$, as stated in the Mermin-Wagner-Hohenberg's theorem \cite{Mermin1966,Hohenberg1967}.
As is clear from this structure, the giant fluctuations are activated by the longitudinal noise $\sigma_{\parallel\parallel}$ that get converted to the Goldstone mode through the non-Hermitian-induced mixing $\zeta=2\kappa$.

It is interesting to compare this result to an $O(N)$ model in a static random field studied by Imry and Ma \cite{Imry1975}, where the correlation function of the transverse magnetization also diverges at $d\le 4$.
In that case, the anomalous fluctuations emerge from coupling between a \textit{static} random field and the Goldstone mode, causing the system to separate into domains.
Our case may be viewed as a \textit{dynamical} extension of this discussion, where the coupling between the longitudinal white noise and the Goldstone mode triggers the anomalous fluctuations.

Before closing this section, we estimate how close to the CEP one should tune the parameters to see the above enhancement of the phase fluctuations.
Let us consider first the conventional critical phenomena, say of a (relaxational) Ising model, where the correlation function in the vicinity of the critical point is given by
\begin{eqnarray}
\label{correlator Ising}
\avg{\delta\phi(\bm k)\delta\phi(-\bm k)}
\sim
\frac{\sigma}{\nu\bm k^2}
\bigg[
1+\frac{m}{\nu\bm k^2}
+O\bigg(
\Big(
\frac{m}{\nu\bm k^2}
\Big)^2
\bigg)
\bigg],
\end{eqnarray}
where $\delta\phi$ is the fluctuations of the Ising spins, $m$ is the damping gap, $\nu$ is the diffusion constant, and $\sigma\sim k_{\rm B}T$ is the thermal noise strength ($T$ is the temperature).
$m$ characterizes the distance from the critical point. 
Here, we have expanded the correlator in terms of $m$, which is justified at the short wavelength regime compared to the correlation length, i.e.
$|\bm k|\gg\xi^{-1}\sim\sqrt{m/\nu}$.
However, the fluctuations are dominated in the long wavelength regime longer than the thermal de Broglie wavelength (which is the wavelength that makes the leading factor of Eq. (\ref{correlator Ising}), $\sigma/(\nu\bm k^2)$, order unity)
$|\bm k|\lesssim \lambda_{\rm T}^{-1}\sim\sqrt{\sigma/\nu}$.
Thus, the enhancement of the correlation is realized when there exists a momentum window that satisfies both, i.e. $  \lambda_{\rm T}\lesssim\xi$, or
\begin{eqnarray}
\label{critical region Ising}
m\lesssim \sigma.
\end{eqnarray}
Physically, this means that the critical fluctuations appear when the damping rate $m$ is small enough for the thermal noise to populate the gapped mode before it damps. 
Note how this critical region scales linearly in terms of the noise strength $\sigma$. 

In the vicinity of the CEP, on the other hand, the correlator Eq. (\ref{fluctuation_CEP}) is expanded in terms of the distance from the CEP $\gamma$ as,
\begin{eqnarray}
\label{correlator vicinity CEP}
\avg{\delta\theta_\alpha(\bm k)\delta\theta_\beta(-\bm k)}
\sim\frac{A}{\bm k^4} 
\bigg[
1+\frac{\gamma}{D\bm k^2}
+O\bigg(\Big(\frac{\gamma}{D\bm k^2}\Big)^2\bigg)
\bigg].
\end{eqnarray}
Therefore, the correlation length $\xi_{\rm CEP}$ (which sets the wavelength that justifies the above expansion) is estimated as $\xi_{\rm CEP}\sim\sqrt{D/\gamma}$, while the effective thermal de Broglie length (that makes $A/\bm k^4 > O(1)$ at $|\bm k|\lesssim (\lambda_{\rm T}^{\rm CEP})^{-1}$) is estimated as $\lambda_{\rm T}^{\rm CEP}\sim A^{-1/4}$.
Following the same logic as above, we can estimate the critical region as $\lambda_{\rm T}^{\rm CEP}\lesssim\xi_{\rm CEP}$, or
\begin{eqnarray}
\label{critical region CEP}
\gamma\lesssim\gamma_c=\frac{\kappa\sqrt{D\sigma_{\parallel\parallel}}}{v}.
\end{eqnarray}
Thus, the critical region scales as $\gamma_c\sim\sqrt{\sigma_{\parallel\parallel}}$, in contrast to the linear scaling for conventional critical points (Eq. (\ref{critical region Ising})).
This difference attributes to the collective mode coalescence to the Goldstone mode at the CEP. 
Since, in the vicinity of the CEP, the gapped mode is almost collinear to the Goldstone mode, it is easier for the noise to excite this mode compared to that in the conventional case.  

\section{Strong-coupling fixed points at spatial dimension $d<d_c=8$}

We now put back the nonlinear terms to analyze the dynamic critical behavior of the CEP.
Following the standard procedure of the dynamic renormalization group method \cite{Hohenberg1977}, we first compute the perturbative correction from the nonlinear couplings and then rescale space, time, and phase fluctuations according to,
\begin{eqnarray}
\bm r\rightarrow e^{l} \bm r, t\rightarrow e^{zl} t,\delta\theta_\alpha\rightarrow e^{\chi_\alpha l}\delta\theta_\alpha,
\label{scaling}
\end{eqnarray}
to formulate the flow equations.
Finding the fixed point of the flow equations provides the universal scaling features of the CEP, such as,
\begin{eqnarray}
\avg{\delta\theta_\alpha(\bm r,t)\delta\theta_\beta(\bm r',t'))}
&=&|\bm r-\bm r'|^{\chi_\alpha+\chi_\beta}f_{\alpha\beta}
\Big(\frac{t-t'}{|\bm r-\bm r'|^z}\Big),
\nonumber\\
\label{phase-phase}
\end{eqnarray}
where $f_{\alpha\beta}(x)$ is a scaling function.

In our two-component case, however, not all of the parameters can be fixed simultaneously under renormalization. For instance, the rescaling (\ref{scaling}) changes $\kappa,v$ and $D$ as $\kappa\rightarrow \kappa e^{zl},v\rightarrow v e^{(z-1)l}$ and $D\rightarrow D e^{(z-2)l}$, respectively,
but when we require, as usual, the lowest-order kinetic term (i.e. the velocity $v$) to be fixed,
$\kappa$ flows to infinity as $l\rightarrow\infty$ while $D\rightarrow0$. 
Here, we require  $A=\kappa^2\sigma_{\parallel\parallel}/(Dv^2),\tilde\gamma\equiv\gamma/D$, and $v$ to be fixed, since the correlation function within the linearized theory is determined solely by these parameters in the vicinity of the CEP in the limit $D\rightarrow 0$, as seen in Eq. (\ref{correlator vicinity CEP}).
We have implicitly assumed here that the strong-coupling fixed point is not very far away from the Gaussian fixed point ($\gamma_*=t_{\alpha*}=\lambda_{\alpha\beta*}^\gamma=0$), which we will justify by restricting ourselves to  spatial dimensions close to the upper critical dimension $d_c$.
The above assumption also requires $\chi_{\rm l}=\chi_{\rm g}(\equiv \chi)$. 
In the linearized theory, we get the roughening exponent $\chi=\chi_{\rm G}=(4-d)/2$ and the dynamic exponent $z=z_{\rm G}=1$.

The flow equations within the one-loop order read, (See the Appendix \ref{appendixRG} for derivation.) 
\begin{eqnarray}
\frac{d\tilde\gamma}{dl}
&=&\bigg[2- \Big(
\frac{C_8}{32d} 
+\frac{C_{10}}{16d}
\tilde\gamma
\Big)\Gamma
\bigg]\tilde\gamma, 
\label{betakappa}
\\
\frac{dv}{dl}
&=&
(z-1)v, 
\label{betav}
\\
\frac{dA}{dl}
&=& \bigg[
4-d-2\chi -  \Big(
\frac{C_8}{32d} 
+\frac{C_{10}}{16d}
\tilde\gamma
\Big)\Gamma 
\bigg]A,
\label{betaA}
\end{eqnarray}
where we have retained only the most relevant coupling. Here, $C_i=(S_d/(2\pi)^d)\Lambda_c^{d-i}$ with $S_d$ the surface area of a $d$-dimensional sphere.
Since $D$ is (dangerously) irrelavant and $\kappa$ flows to infinity, the most relevant coupling is the term that have the lowest order on $D$ and the highest order on $\kappa$,  which turns out to be,
\begin{eqnarray}
\Gamma
& \equiv&\frac{t_{\parallel}\sigma_{\parallel\parallel}} {D^5}
(t_{\parallel} v^2 +4\kappa^2\lambda^\parallel_{\perp\perp}). 
\label{Gamma}
\end{eqnarray} 
Here, $t_\parallel=\sqrt{2}(t_{\rm l}-t_{\rm g})$ is the massive out-of-phase nonlinearity 
and 
$\lambda^\parallel_{\perp\perp}=[(\lambda^{\rm l}_{\rm ll}+\lambda^{\rm l}_{\rm lg}+\lambda^{\rm l}_{\rm gg})
-(\lambda^{\rm g}_{\rm ll}+\lambda^{\rm g}_{\rm lg}+\lambda^{\rm g}_{\rm gg})]/(2\sqrt{2})$ 
is the KPZ-like nonlinear coupling that converts the two incoming Goldstone mode into the longitudinal mode. 
The effective coupling $\Gamma$ follows the flow equation,
\begin{eqnarray}
\frac{d\Gamma}{dl}
&=&\bigg[
8-d - \Big(
\frac{5C_8}{32d} 
+\frac{5C_{10}}{16d}
\tilde\gamma
\Big)\Gamma\bigg]\Gamma,
\label{betaGamma}
\end{eqnarray}
indicating that $\Gamma$ is relevant at dimensions as high as $d\le d_c=8$, which is to be compared to the conventional critical point in the Ising model with $d_c=4$ \cite{Altland_Simons} or the KPZ scaling with $d_c=2$ \cite{Kardar1986}.

We remark that the upper critical dimension $d_c=8$,
which characterizes the strength of  many-body correlation effects,
is higher than that obtained from the trivial scaling, which are $d_c=1$ and $d_c=3$ for $\lambda^\parallel_{\perp\perp}$ and $t_\parallel$, respectively. 
This anomalously enhanced many-body correlation effect is again due to the coalescence of collective eigenmodes to the Goldstone mode. 
As we have discussed earlier, 
the mode coalescence leads  to the appearance of the linear dispersion at the CEP (i.e. the term $\pm v|\bm k|$ in  Eq. (\ref{GoldstoneCEP})). 
This makes the diffusion term $-i D\bm k^2$ dangerously irrelevant, 
 i.e. $D\rightarrow 0$,
which makes the system infinitely sensitive to 
the noise that activates the fluctuations, see Eq. (\ref{def A}).
This singular sensitivity turns the effective nonlinear coupling $\Gamma$ relevant even at dimensions where the nonlinear couplings $\lambda^\parallel_{\perp\perp}$ and $t_\parallel$ themselves are irrelevant, because the factor $D^5$ in the denominator of Eq. (\ref{Gamma}) also flows to zero as approaching the fixed point.


At spatial dimensions close to the upper critical dimension, $d=d_c-\epsilon=8-\epsilon$, we find a strong-coupling fixed point at $(\tilde\gamma_*,\Gamma_*)\approx(0, \epsilon(32\cdot 8)/(5C_8))$, associated with a new universality class with critical exponents,
\begin{eqnarray}
\chi&\approx&\chi_{\rm G}
-\frac{\epsilon}{10}, 
z=z_{\rm G}. 
\end{eqnarray}
We briefly note that, as a consequence of the two-component nature, $\Gamma$ may take either sign.   
Since $\Gamma$ cannot change its sign during the flow because $d\Gamma/dl=0$ at $\Gamma=0$ as seen in Eq. (\ref{betaGamma}), the system can only flow to the above-obtained strong-coupling fixed point when the bare parameter $\Gamma(l=0)$ is positive, which is determined by the balance between the diffusion rate of the pump and the nonlinear loss rate of the field (See Eq. (\ref{v2_supp}).).
Otherwise, when $\Gamma(l=0)<0$, the flow direct towards $\Gamma\rightarrow -\infty$, implying the existence of another distinct phase of matter, which will be explored in the future work. 

As a final remark, we ask: how close to the CEP should we be for the nonlinear many-body correlation effect to be detectable? 
We attempt to answer this by observing that our CEP flow equations  (\ref{betakappa})-(\ref{betaGamma}) are derived under the assumption that we are in the vicinity of the CEP, where we can expand the valuables in terms of $\tilde\gamma(l)$. 
This is only justified when $\tilde\gamma(l)\cdot \Lambda_c^{-2}<1$.

Firstly, if the bare quantities is in the regime $\tilde\gamma(l=0) \Lambda_c^{-2}>1$, since Eq. (\ref{betakappa}) makes $\tilde\gamma(l)$ to be monotonically increasing function in terms of $l$, there is no region that is appropriate to use our CEP flow equations (\ref{betakappa})-(\ref{betaGamma}).
Thus, we would not see any CEP physics in this regime. 
This is consistent with the critical region (\ref{critical region CEP}) estimated in Sec. \ref{sec linear}, when we estimate the ultraviolet cutoff to be $\Lambda_c\sim(\lambda^{\rm CEP}_{\rm T})^{-1}\sim A^{1/4}$.  This is reasonable because the shortest length scale that enters the physics in this semiclassical treatment is the shortest length scale of fluctuations occupied by the white noise, which is the effective de Broglie length $\lambda^{\rm CEP}_{\rm T}$.

Next, we consider the critical region $\tilde\gamma(l=0)\Lambda_c^{-2}<1$. In this case, our CEP flow equations (\ref{betakappa})-(\ref{betaGamma}) are valid in the initial flow but stops being justified at $l=l_0$ where $\tilde\gamma(l)$ has grown to 
$\tilde\gamma(l_0)\Lambda_c=1$. 
Ignoring $\Gamma(l)$ for simplicity, we get 
$\tilde\gamma(l)\simeq e^{2l}\tilde\gamma(l=0)$, which gives
\begin{eqnarray}
e^{l_0}=[\tilde\gamma(l=0)]^{-\frac{1}{2}}\Lambda_c.
\end{eqnarray}
In the case where $\Gamma(l)\Lambda_c^{d-8}\lesssim 1$ in the CEP region $l=[0,l_0]$, the nonlinear many-body correlation effects can be safely neglected but otherwise the non-Gaussian fluctuations potentially get dominated. 
The latter region can be estimated as 
\begin{eqnarray}
\Gamma(l_0)\Lambda_c^{d-8}
&\simeq & 
\Lambda_c^{d-8}  e^{(8-d)l_0}\Gamma(l=0) 
\nonumber\\
&\simeq& 
\Gamma(l=0) 
[\tilde\gamma(l=0)]^{-\frac{8-d}{2}}
\gtrsim 1,
\end{eqnarray}
or
\begin{eqnarray}
\label{non Gaussian}
\gamma
\lesssim 
D\Gamma^{\frac{2}{8-d}}
\end{eqnarray}
in terms of the bare quantities. 
Thus, the critical region where the non-Gaussian fluctuations can be detected is given by the regions where the bare quantities satisfies both Eqs. (\ref{critical region CEP}) and (\ref{non Gaussian}). 
Notice the low power on the left-hand side that makes it easier to get into the non-Gaussian regimes with small nonlinearities, which is to be compared to the conventional Ising model case where this power is replaced by $2/(4-d)$ \cite{Amit1973}.
This is originated from the anomalously high upper critical dimension $d_c=8$.



\section{Summary}
\label{sec_summary}

To summarize, we have proposed a novel mechanism for the occurance of the dynamic critical phenomena that arise at the CEP, originated from the coalescence of the collective eigenmodes to the Goldstone mode. 
We showed that this peculiar property gives rise to anomalously giant phase fluctuations that diverge at $d\le 4$. 
It also leads to the appearance of sound mode that anomalously enhance the many-body correlation effects, which survive at exceptionally high spatial dimensions ($d<8$). 

Of course, our $\epsilon(=8-d)$-expansion performed in this study cannot be directly applied to the physical dimensions $d=1,2,3$. 
Direct numerical simulation of the stochastic GP equation (\ref{GP}) at the CEP to extract the critical exponents at these dimensions $d=1,2,3$ is under progress. 
However, our analysis clearly demonstrates the key aspect of this novel dynamical critical phenomena at the CEP, which is the anomalous enhancement of the many-body correlation effects, where we find a stable strong-coupling fixed point just below the upper critical dimension $d_c=8$ associated with a new universality class beyond the classification by Hohenberg and Halperin \cite{Hohenberg1977}. 
We expect the dynamic scaling law to be different from any known scaling in the physical dimensions as well since the origin of the criticality is fundamentally different.
The signatures of these anomalous critical fluctuations at the CEP should be observable through the measurement of the phase-phase correlator (Eq. (\ref{phase-phase})), which is experimentally accessible in e.g. polariton condensates using the Michelson interferometer \cite{Caputo2018}.
We also believe an anomalous signal would appear in various response functions as well, although we have not studied them thoroughly, remaining as our future work. 

In this work, we have implicity restricted ourselves to situations where the amplitude of the macroscopic wave funtions in the steady state $|\Psi_{\rm l}^0|, |\Psi_{\rm g}^0|$ is large enough such that the amplitude fluctuations dynamics are overdamped. 
It is interesting to ask how the critical fluctuations would change their character when this restriction is lifted. 

We thank S. Diehl, V. Vitelli, M. Fruchart, and A. Edelman for critical reading of the manuscript and giving helpful comments. We acknowledge A. Galda for fruitful discussions. R.H. was supported by a Grand-in-Aid for JSPS fellows (Grant No. 17J01238).  Work at Argonne National Laboratory is supported by the U. S. Department of Energy, Office of Science, BES-MSE under Contract No. DE-AC02-06CH11357.

\begin{appendix}

\begin{widetext}

\section{Steady state solution}
\label{appendix NESS}

Unless dynamically unstable, the coupled driven dissipative Gross-Pitaevskii (GP) equation (\ref{GP}) approaches a steady state at long-time limit $t\rightarrow\infty$ due to the nonlinear saturation term ($v_{\rm g}|\Phi_{\rm g}^0|^2$). 
In the absense of noise, the steady state is described by the ansatz, $\Psi_\alpha^0(t)=\Phi^0_\alpha e^{-iEt}$, which reduces Eq. (\ref{GP}) to the steady state condition for $\Phi_\alpha^0$, 
\begin{eqnarray}
E\left(\begin{array}{c}
\Phi^0_{\rm l} \\
\Phi^0_{\rm g}  
\end{array}\right)
&=&
A_{\rm GP}(\Phi_{\rm l}^0,\Phi_{\rm g}^0)
\left(\begin{array}{c}
\Phi^0_{\rm l} \\
\Phi^0_{\rm g}  
\end{array}\right)
=
\left(\begin{array}{cc}
\omega_{\rm l}-i\kappa & g \\
g & \omega_{\rm g} 
+
U_{\rm g} |\Phi^0_{\rm g}|^2
+i(P - v_{\rm g}|\Phi^0_{\rm g}|^2)
\end{array}\right)
\left(\begin{array}{c}
\Phi^0_{\rm l} \\
\Phi^0_{\rm g}  
\end{array}\right).
\label{GP_steady_supp}
\end{eqnarray}
Here, $E$ is the emission energy of the condensate that needs to be real such that it is consistent with our assumption that $\Phi_\alpha^0$ a steady state solution. 
Diagonalizing this eigenvalue problem gives
\begin{eqnarray}
\label{diagonal AGP}
E
\left(\begin{array}{c}
\Phi^0_{-} \\
\Phi^0_{+}  
\end{array}\right)
=
\left(\begin{array}{cc}
E_- &  0 \\
0 & E_+
\end{array}\right)
\left(\begin{array}{c}
\Phi^0_{-} \\
\Phi^0_{+}  
\end{array}\right),
\end{eqnarray}
where the order parameter is transformed according to
\begin{eqnarray}
\left(\begin{array}{c}
\Phi^0_{-} \\
\Phi^0_{+}  
\end{array}\right)
=
U
\left(\begin{array}{c}
\Phi^0_{\rm l} \\
\Phi^0_{\rm g}  
\end{array}\right),
\end{eqnarray}
with $U^{-1}=(\bm u_-,\bm u_+)$.
Here,
\begin{eqnarray}
E_\pm&=&\frac{1}{2}
\Big[
\omega_{\rm l}+\omega_{\rm g}+U_{\rm g}|\Phi_{\rm g}^0|^2-i(\kappa-P+v_{\rm g}|\Phi_{\rm g}^0|^2)
\pm \Omega
\Big],\\
\Omega &=&
\sqrt{
\tilde\delta^2+4g^2
-(\kappa+P-v_{\rm g}|\Phi_{\rm g}^0|^2)^2
-2i\tilde\delta(\kappa+P-v_{\rm g}|\Phi_{\rm g}^0|^2)
}
\end{eqnarray}
are the eigenvalues of $\hat A_{\rm GP}$ with the corresponding eigenvectors given by, 
\begin{eqnarray}
\bm u_-=
\left(\begin{array}{c}
\frac{\Omega-\varphi}{2} \\
-g
\end{array}\right), 
	\qquad
\bm u_+=
\left(\begin{array}{c}
g \\
\frac{\Omega-\varphi}{2}
\end{array}\right),
\label{eigenvectors A} 
\end{eqnarray}
for $E_-$ and $E_+$, respectively.
We have introduced the complex detuning
\begin{eqnarray}
\varphi=\omega_{\rm l}-\omega_{\rm g}-U_{\rm g}|\Phi_{\rm g}^0|^2-i(\kappa+P-v_{\rm g}|\Phi_{\rm g}^0|^2)
\end{eqnarray}
and the effective detuning $\tilde \delta={\rm Re}\varphi$.

Equation (\ref{diagonal AGP}) shows that the steady state solution can be classified into two types \cite{Hanai2019} (other than the trivial solution $\Phi_-^0=\Phi_+^0=0$), namely, the ``$-$'' solution with $(E=E_-, \Phi_-^0\ne 0, \Phi_+^0=0)$ and the ``$+$'' solution with $(E=E_+, \Phi_+^0\ne 0, \Phi_-^0=0)$.
These solutions correspond to a macroscopically occupied state into the lower and upper branch, respectively. 
This discussion shows that, unlike in similar non-Hermitian systems where the transient dynamics is discussed, a superposition of the two eigenstates is forbidden due to the steady state constraint.
This is reasonable since the simultaneous occupation of two eigenmodes will cause beating, which is obviously not a steady state. 

Crucially, the matrix $A_{\rm GP}(\Phi_{\rm l}^0,\Phi_{\rm g}^0)$ is a non-Hermitian matrix, which makes the eigenvectors $\bm u_\pm$ (Eq. (\ref{eigenvectors A})) non-orthorgonal, giving rise to an exceptional point where they become collinear. 
This means that the two solution types coalesce at this point, in quite the same manner to the critical point of the liquid-gas phase diagram where the liquid and gas solutions coalesce at that point.
As proven in Ref. \cite{Hanai2019}, this point indeed marks the endpoint of a first-order-like phase boundary between the two types of solutions (See Fig. \ref{fig_phasediagram}). 
Since this point exhibits critical fluctuations as shown in the main text, which we dub the critical exceptional point (CEP).

The CEP is found at the parameter that satisfies 
\begin{eqnarray}
E=E_- = E_+.
\end{eqnarray}
For this to be true, $\Omega = 0$ needs to be satisfied. 
In such case, the steady state constraint requires $E$ to be real, giving 
\begin{eqnarray}
\label{Phig_EP}
|\Phi_{\rm g}^0|^2= \frac{P-\kappa}{v_{\rm g}},
\end{eqnarray}
which sets the amplitude of the gain component $|\Phi_{\rm g}^0|$ in the steady state. 
It is clear from this expression that the pump power needs to be $P>\kappa$ such that we realize the CEP. 

To obtain the condition to realize the CEP, we look for parameters that makes $\Omega$ vanish. 
For the imaginary part of $\Omega^2$ to vanish, we need to set to an effective zero detuning,  
\begin{eqnarray}
\label{on resonance}
\tilde\delta = \omega_{\rm l}-\omega_{\rm g}-U_{\rm g}|\Phi_{\rm g}^0|^2 
= \omega_{\rm l}-\omega_{\rm g}-\frac{(P-\kappa)U_{\rm g}}{v_{\rm g}} =0
\end{eqnarray}
Substituting Eqs. (\ref{Phig_EP}) and (\ref{on resonance}) back to $\Omega$, we get the condition  
\begin{eqnarray}
\kappa = g.
\label{kappa EP}
\end{eqnarray}

We emphasize that, these conditions requires only \textit{two} parameters to fine-tune, i.e. conditions (\ref{on resonance}) and (\ref{kappa EP}), where Eq. (\ref{Phig_EP}) is automatically satisfied in the steady state at the CEP. 

\section{Noisy driven-dissipative Gross-Pitaevskii equation to the Kardar-Parisi-Zhang equation}
\label{appendixKPZ}

Here, we derive the Kardar-Parisi-Zhang (KPZ)-like stochastic equation of motion (\ref{coupledKPZ}) from the coupled driven-dissipative GP equation, 
\begin{eqnarray}
i\partial_t 
\left(\begin{array}{c}
\Psi_{\rm l}(\bm r,t) \\
\Psi_{\rm g}(\bm r,t) 
\end{array}\right)
&=&\left(\begin{array}{cc}
\omega_{\rm l}-i\kappa- K_{\rm l}\nabla^2 & g \\
g & \omega_{\rm g} +iP -(K_{\rm g}-iD_{\rm g})\nabla^2+(U_{\rm g}-iv_{\rm g})|\Psi_{\rm g}(\bm r,t)|^2
\end{array}\right)
\left(\begin{array}{c}
\Psi_{\rm l}(\bm r,t) \\
\Psi_{\rm g}(\bm r,t) 
\end{array}\right)
+\left(\begin{array}{c}
\eta_{\rm l}(\bm r,t) \\
\eta_{\rm g}(\bm r,t) 
\end{array}\right).
\nonumber\\
\label{GP_supp}
\end{eqnarray}
We rewrite the GP equations (\ref{GP_supp}) in terms of the amplitude and phase fluctuations around the uniform steady state. Here, we define the amplitude and phase fluctuations by $\Psi_\alpha(\bm r,t)=[\Phi^0_\alpha+\delta\Phi_\alpha(\bm r,t)]e^{-iEt}
=(M^0_\alpha+\delta|\Phi_\alpha(\bm r,t)|)e^{i(\theta^0_\alpha+\delta\theta_\alpha(\bm r,t))} e^{-iEt}$, where $\Phi_\alpha^0=M^0_\alpha e^{i\theta^0_\alpha}$.
Assuming that the amplitude fluctuations are small such that the terms with $O((\delta|\Phi_\alpha|)^2)$ are negligible, we obtain,
\begin{eqnarray}
&&-M_{\rm l}^0\partial_t\delta\theta_{\rm l}(\bm r,t)
= (\omega_{\rm l}-E)\delta|\Phi_{\rm l}(\bm r,t)|
+ g M_{\rm g}^0 [\cos(\Delta\theta_0+\delta\theta_{\rm l}(\bm r,t)-\delta\theta_{\rm g}(\bm r,t)) - \cos\Delta\theta_0]
\nonumber\\
&& \ \ \ \ \ \ \ \ \ \ \ \ \ \ \ \ \ \ \ \ 
+g\cos(\Delta\theta_0+\delta\theta_{\rm l}(\bm r,t)-\delta\theta_{\rm g}(\bm r,t))\delta|\Phi_{\rm g}(\bm r,t)|
+K_{\rm l}M_{\rm l}^0(\nabla\theta_{\rm l}(\bm r,t))^2
+\eta_{\rm l}'(\bm r,t),
\label{GP_Re1_supp} \\
&&\partial_t\delta|\Phi_{\rm l}(\bm r,t)|
=-\kappa\delta|\Phi_{\rm l}(\bm r,t)|
-K_{\rm l}M_{\rm g}^0\nabla^2\theta_{\rm l}(\bm r,t)
- g M_{\rm g}^0 [\sin(\Delta\theta_0+\delta\theta_{\rm l}(\bm r,t)-\delta\theta_{\rm g}(\bm r,t)) - \sin\Delta\theta_0]
\nonumber\\
&& \ \ \ \ \ \ \ \ \ \ \ \ \ \ \ \ 
-g\sin(\Delta\theta_0+\delta\theta_{\rm l}(\bm r,t)-\delta\theta_{\rm g}(\bm r,t))\delta|\Phi_{\rm g}(\bm r,t)|
+\eta_{\rm l}''(\bm r,t), 
\label{GP_Im1_supp} \\
&&-M_{\rm g}^0\partial_t\delta\theta_{\rm g}(\bm r,t)
=
g M_{\rm l}^0 [\cos(\Delta\theta_0+\delta\theta_{\rm l}(\bm r,t)-\delta\theta_{\rm g}(\bm r,t)) - \cos\Delta\theta_0]
+g\cos(\Delta\theta_0+\delta\theta_{\rm l}(\bm r,t)-\delta\theta_{\rm g}(\bm r,t))\delta|\Phi_{\rm l}(\bm r,t)|
\nonumber\\
&& \ \ \ \ \ \ \ \ \ \ \ \ \ \ \ \ \ \ \ \ \ 
+(\omega_{\rm g}+3U_{\rm g}|\Phi_{\rm g}^0|^2-E)\delta|\Phi_{\rm g}(\bm r,t)|
-D_{\rm g}M_{\rm g}^0 \nabla^2\theta_{\rm g}(\bm r,t)
+K_{\rm g}M_{\rm g}^0(\nabla\theta_{\rm g}(\bm r,t))^2
+\eta_{\rm g}'(\bm r,t),
\label{GP_Re2_supp} \\
&&\partial_t\delta|\Phi_{\rm g}(\bm r,t)|
=
-K_{\rm g}M_{\rm g}^0\nabla^2\theta_{\rm g}(\bm r,t)
+ g M_{\rm l}^0 [\sin(\Delta\theta_0+\delta\theta_{\rm l}(\bm r,t)-\delta\theta_{\rm g}(\bm r,t)) - \sin\Delta\theta_0]
\nonumber\\
&& \ \ \ \ 
+g\sin(\Delta\theta_0+\delta\theta_{\rm l}(\bm r,t)-\delta\theta_{\rm g}(\bm r,t))\delta|\Phi_{\rm l}(\bm r,t)|
+(P-3v_{\rm g}|\Phi_{\rm g}^0|^2)\delta|\Phi_{\rm g}(\bm r,t)|
-D_{\rm g}M_{\rm g}^0(\nabla\theta_{\rm g}(\bm r,t))^2
+\eta_{\rm g}''(\bm r,t), 
\label{GP_Im2_supp}
\end{eqnarray}
where $\Delta\theta_0=\theta_{\rm l}^0-\theta_{\rm g}^0$, and $\eta_\alpha(\bm r,t)=\eta_\alpha'(\bm r,t)+i\eta_\alpha''(\bm r,t)$.
Following Ref. \cite{Altman2015}, we further assume that the amplitude fluctuations are overdamped 
(i.e. we neglect the time-derivative in the left-hand side of Eqs. (\ref{GP_Im1_supp}) and (\ref{GP_Im2_supp})).
This brings us to the desired form,
\begin{eqnarray}
\partial_t\delta\theta_\alpha(\bm r,t) = 
\sum_{\beta={\rm l,g}}
W_{\alpha\beta}(\nabla)
\delta\theta_\beta(\bm r,t)
+ t_\alpha (\delta\theta_{\rm l}(\bm r,t)-\delta\theta_{\rm g}(\bm r,t))^2
+\sum_{\beta,\gamma={\rm l,g}}
\lambda^\alpha_{\beta\gamma}
(\nabla\delta\theta_\beta(\bm r,t))
\cdot
(\nabla\delta\theta_\gamma(\bm r,t))
+\xi_\alpha(\bm r,t),
\label{coupledKPZ_supp}
\end{eqnarray} 
with
\begin{eqnarray*}
W(\nabla)
=
\left(\begin{array}{cc}
-s_{\rm l}+\nu_{\rm ll}\nabla^2 
& s_{\rm l}+\nu_{\rm lg}\nabla^2 \\
-s_{\rm g}+\nu_{\rm gl}\nabla^2 
& s_{\rm g}+\nu_{\rm gg}\nabla^2
\end{array}\right),
\end{eqnarray*}
where we have neglected the higher-order massive term $O ((\delta\theta_{\rm l}-\delta\theta_{\rm g})^3)$ that does not contribute to the critical properties within the one-loop order. 

To obtain the explicit expression of the parameters in the coupled KPZ-like equations (\ref{coupledKPZ_supp}) in terms of those in the coupled driven-dissipative GP equations (\ref{GP_supp}) (which is crucial for determining the stability condition), we firstly need to solve the steady state condition Eq. (\ref{GP_steady_supp}).
This requires numerical computation in general, but for the special case where the ``effective detuning'' $\tilde\delta=\omega_{\rm l}-\omega_{\rm g}-U_{\rm g}|\Phi_{\rm g}^0|^2$ is ``on resonance'', $\tilde\delta=0$,  an analytic form can be obtained. This includes the critical exceptional point (CEP), our main focus of this paper, which is found at $\tilde\delta=0$ and $\kappa=g=P-v_{\rm g}|\Phi_{\rm g}^0|^2$. (See Appendix \ref{appendix NESS}.)

Plugging these into Eqs. (\ref{GP_Re1_supp})-(\ref{GP_Im2_supp}), we obtain the explicit form of the parameters in Eq. (\ref{coupledKPZ_supp}).
Here, we provide the list of the parameters for the ``$-$'' solution:
\begin{eqnarray}
&&
s_{\rm l}=\frac{g^2}{\kappa},
s_{\rm g}
=2\kappa-\frac{g^2}{\kappa}
-\frac{U_{\rm g}\sqrt{g^2-\kappa^2}}{v_{\rm g}}, 
\\
&&
\nu_{\rm ll}=\frac{-\sqrt{g^2-\kappa^2}(\kappa-2v_{\rm g} \kappa |\Phi_{\rm g}^0|^2)
K_{\rm l}}
{2v_{\rm g} \kappa |\Phi_{\rm g}^0|^2},
\nu_{\rm lg}=\frac{\sqrt{g^2-\kappa^2}K_{\rm l}}
{2v_{\rm g} |\Phi_{\rm g}^0|^2},
\nonumber\\
&&
\nu_{\rm gl}=-\frac{K_{\rm l}(U_{\rm g}\kappa+v_{\rm g}\sqrt{g^2-\kappa^2})}{v_{\rm g}\kappa},
\nu_{\rm gg}
=D_{\rm g}+\frac{K_{\rm g}U_{\rm g}}{v_{\rm g}}
+\frac{(K_{\rm g}-K_{\rm l})\sqrt{g^2-\kappa^2}}{2v_{\rm g}|\Phi_{\rm g}|^2}
+\frac{K_{\rm l}\kappa \sqrt{g^2-\kappa^2}}{2 g^2 v_{\rm g}|\Phi_{\rm g}|^2}
\Big(
\kappa-\frac{U_{\rm g}\sqrt{g^2-\kappa^2}}{v_{\rm g}}
\Big), 
\nonumber\\
\\
&&
\lambda^{\rm l}_{\rm ll}
=\frac{K_{\rm l}(\kappa-2v_{\rm g}|\Phi_{\rm g}|^2)
(g^2-\kappa(\kappa-2v_{\rm g}|\Phi_{\rm g}|^2))}
{2v_{\rm g}^2 \kappa |\Phi_{\rm g}|^2 }, 
\lambda^{\rm l}_{\rm lg}
=\frac{K_{\rm l}(-\kappa+v_{\rm g}|\Phi_{\rm g}|^2)
(g^2-\kappa(\kappa-2v_{\rm g}|\Phi_{\rm g}|^2))}
{v_{\rm g}^2 \kappa |\Phi_{\rm g}|^2 }, \nonumber\\
&&
\lambda^{\rm l}_{\rm gg}=
\frac{K_{\rm l}(g^2-\kappa^2+2v_{\rm g}\kappa|\Phi_{\rm g}|^2-2v_{\rm g}^2|\Phi_{\rm g}|^4)}{2v_{\rm g}^2|\Phi_{\rm g}|^4},
\nonumber\\
&&
\lambda^{\rm g}_{\rm ll}
=\frac
{1}{2g^2v_{\rm g}^3\kappa |\Phi_{\rm g}^0|^4}
\Big[
g^4 v_{\rm g}[(K_{\rm g}-K_{\rm l})\kappa
+2K_{\rm l}v_{\rm g}|\Phi_{\rm g}^0|^2 ]
+g^2\kappa
[(2K_{\rm l}-K_{\rm g})v_{\rm g}\kappa^2 
-K_{\rm l}U_{\rm g}\kappa\sqrt{g^2-\kappa^2}
\nonumber\\
&& \ \ \ \ 
+2(-3K_{\rm l}+K_{\rm g})v_{\rm g}^2\kappa |\Phi_{\rm g}^0|^2
+2K_{\rm l}U_{\rm g}v_{\rm g}\sqrt{g^2-\kappa^2}|\Phi_{\rm g}^0|^2
+2K_{\rm l}v_{\rm g}^3|\Phi_{\rm g}^0|^4
]
\nonumber\\
&&\ \ \ \ 
-K_{\rm l}\kappa^2
(v_{\rm g}\kappa-U_{\rm g}\sqrt{g^2-\kappa^2}|\Phi_{\rm g}^0|^2)(\kappa^2
- 4v_{\rm g}\kappa |\Phi_{\rm g}^0|^2
+2v_{\rm g}^2|\Phi_{\rm g}^0|^4)
\Big], 
\nonumber\\ 
&&
\lambda^{\rm g}_{\rm lg}
=-\frac
{1}{2g^2v_{\rm g}^3\kappa |\Phi_{\rm g}^0|^4}
\Big[
K_{\rm l}[g^2v_{\rm g} + \kappa(-v_{\rm g}\kappa+U_{\rm g}\sqrt{g^2-\kappa^2})]
(-\kappa+v_{\rm g} |\Phi_{\rm g}^0|^2)
(g^2-\kappa(\kappa-2v_{\rm g} |\Phi_{\rm g}^0|^2))
\nonumber\\
&&\ \ \ \ 
+g^2 v_{\rm g}\kappa
[g^2K_{\rm g}+D_{\rm g}v_{\rm g}\sqrt{g^2-\kappa^2} |\Phi_{\rm g}^0|^2
-K_{\rm g}\kappa
(\kappa-2v_{\rm g} |\Phi_{\rm g}^0|^2) ]
\Big],
\nonumber\\
&&\lambda^{\rm g}_{\rm gg}
=\frac{1}{2v_{\rm g}|\Phi_{\rm g}^0|^4}
\Big[
K_{\rm g}(g^2-\kappa^2)+2K_{\rm g}v_{\rm g}\kappa  |\Phi_{\rm g}^0|^2
+2D_{\rm g}v_{\rm g}\sqrt{g^2-\kappa^2} |\Phi_{\rm g}^0|^2
+2 D_{\rm g}U_{\rm g}v_{\rm g} |\Phi_{\rm g}^0|^4
-2K_{\rm g}v_{\rm g}^2 |\Phi_{\rm g}^0|^4
\nonumber\\
&&\ \ \ \ 
+K_{\rm l}
\Big(-1 + \frac{\kappa(\kappa v_{\rm g}-U_{\rm g}\sqrt{g^2-\kappa^2}}
{v_{\rm g} g^2}
\Big)
(g^2-\kappa^2+2v_{\rm g}\kappa  |\Phi_{\rm g}^0|^2
-2v_{\rm g}^2 |\Phi_{\rm g}^0|^4)
\Big],
\\
&&
t_{\rm l}=\frac{g^2\sqrt{g^2-\kappa^2}}{2v_{\rm g}\kappa |\Phi_{\rm g}^0|^2},
t_{\rm g}=\frac
{2\kappa(U_{\rm g}\kappa + v_{\rm g}\sqrt{g^2-\kappa^2})
(\kappa-v_{\rm g} |\Phi_{\rm g}^0|^2)
-g^2(2U_{\rm g}\kappa+v_{\rm g}\sqrt{g^2-\kappa^2})}
{2v_{\rm g}^2\kappa |\Phi_{\rm g}^0|^2},
\\
&&
\xi_{\rm l}(\bm r,t)=\frac{g}{\kappa}\eta_{\rm l}''(\bm r,t),
\nonumber\\
&&
\xi_{\rm g}(\bm r,t)
= \frac{1}{g \kappa v_{\rm g}|\Phi_{\rm g}^0|^2}
\Big[
(\kappa^2 U_{\rm g}+v_{\rm g}\sqrt{g^2-\kappa^2})
\eta'_{\rm l}(\bm r,t)
-[g^2 v_{\rm g}+\kappa(-v_{\rm g}\kappa+U_{\rm g}\sqrt{g^2-\kappa^2})]\eta''_{\rm l}(\bm r,t)
\nonumber\\
&& \ \ \ \ \ \ \ \ \ \ 
-g\kappa v_{\rm g} \eta_{\rm g}'(\bm r,t)
-g\kappa U_{\rm g} \eta_{\rm g}'(\bm r,t)
\Big].
\end{eqnarray} 
Especially at the CEP $g=\kappa$, we find,
\begin{eqnarray}
&&s_{\rm l}=s_{\rm g}=\kappa, \\ 
&&
\nu_{\rm ll}=\nu_{\rm lg}=0,
\nu_{\rm gl}=-\frac{K_{\rm l}U_{\rm g}}{v_{\rm g}},
\nu_{\rm gg}=D_{\rm g}+\frac{K_{\rm g}U_{\rm g}}{v_{\rm g}}, 
\\
&&\lambda^{\rm l}_{\rm ll}=-K_{\rm l}
\Big(-2+\frac{\kappa}{v_{\rm g} |\Phi_{\rm g}^0|^2}\Big), 
\lambda^{\rm l}_{\rm lg}=2K_{\rm l}
\Big(1+\frac{\kappa}{v_{\rm g} |\Phi_{\rm g}^0|^2}\Big),
\lambda^{\rm l}_{\rm gg}=K_{\rm l}
\Big(-1+\frac{\kappa}{v_{\rm g} |\Phi_{\rm g}^0|^2}\Big),
\nonumber\\
&&\lambda^{\rm g}_{\rm ll}
=\frac{K_{\rm g}\kappa}{v_{\rm g} |\Phi_{\rm g}^0|^2} , 
\lambda^{\rm g}_{\rm lg}
=-\frac{2K_{\rm g}\kappa}{v_{\rm g} |\Phi_{\rm g}^0|^2} ,
\lambda^{\rm g}_{\rm gg}
=\frac{D_{\rm g}U_{\rm g}}{v_{\rm g}}
+K_{\rm g}\Big(
-1+\frac{\kappa}{v_{\rm g} |\Phi_{\rm g}^0|^2}
\Big),
\\
&&
t_{\rm l}=0,
t_{\rm g}=-\frac{U_{\rm g}\kappa}{v_{\rm g}},
\\
&&
\xi_{\rm l}(\bm r,t)=\frac{\eta'_{\rm l}(\bm r,t)}{|\Phi_{\rm l}^0|},
\xi_{\rm g}(\bm r,t)
= \frac{-v_{\rm g} \eta'_{\rm g}(\bm r,t)+ U_{\rm g}(\eta''_{\rm l}(\bm r,t)-\eta''_{\rm g}(\bm r,t))}{ v_{\rm g}|\Phi_{\rm g}^0|}.
\end{eqnarray} 
As discussed in the main text, since the collective mode is given by (Eq. (2)), 
\begin{eqnarray}
\omega_\pm (\bm k)=
\frac{1}{2}\big[
-i(\gamma + 2D\bm k^2)\pm \sqrt{-\gamma^2+4v^2 \bm k^2}
\big],
\end{eqnarray}
the stability condition can be determined by $\gamma=s_{\rm l}-s_{\rm g},v^2=[2(s_{\rm g}\nu_{\rm lg}+s_{\rm l}\nu_{\rm gl})+(s_{\rm l}+s_{\rm g})(\nu_{\rm gg}-\nu_{\rm ll})]/2,D=(\nu_{\rm ll}+\nu_{\rm gg})/2\ge 0$.
For the above-obtained ``$-$'' solution, we get,
\begin{eqnarray}
\gamma &=& \frac{2g^2}{\kappa}
-2\kappa + \frac{2U_{\rm g}\sqrt{g^2-\kappa^2}}{v_{\rm g}}
\ge  0, \\
v^2&=& 
\frac{1}{2g^2\kappa^2 v_{\rm g}^3|\Phi_{\rm g}^0|^2}
\bigg[
K_{\rm l}\kappa^5 \big[
2U_{\rm g}v_{\rm g}\kappa - U_{\rm g}^2\sqrt{g^2-\kappa^2}
+v_{\rm g}^2\sqrt{g^2-\kappa^2}
\big]
\nonumber\\
&-&g^4v_{\rm g}\big[
K_{\rm g}\kappa (U_{\rm g}\kappa + v_{\rm g}\sqrt{g^2-\kappa^2})
+K_{\rm l}(U_{\rm g}\kappa^2+2v_{\rm g}^2\sqrt{g^2-\kappa^2} |\Phi_{\rm g}^0|^2)
\big]
\nonumber\\
&+&g^2\kappa^2
\Big[
K_{\rm l}[U_{\rm g}^2\kappa\sqrt{g^2-\kappa^2}
+v_{\rm g}^2\sqrt{g^2-\kappa^2}(\kappa-2v_{\rm g}|\Phi_{\rm g}^0|^2)
-U_{\rm g}v_{\rm g}\kappa
(\kappa+2v_{\rm g}|\Phi_{\rm g}^0|^2)
]\nonumber\\
&+&
v_{\rm g}
\big[
2D_{\rm g}v_{\rm g}
(v_{\rm g}\kappa
-U_{\rm g}\sqrt{g^2-\kappa^2})|\Phi_{\rm g}^0|^2
+K_{\rm g}
(
v_{\rm g}\kappa\sqrt{g^2-\kappa^2}-2U_{\rm g}^2\sqrt{g^2-\kappa^2}|\Phi_{\rm g}^0|^2
+U_{\rm g}\kappa(\kappa+2v_{\rm g})|\Phi_{\rm g}^0|^2
)
\big]
\Big]
\bigg],
\nonumber\\
\\
D&=& \frac{1}{4g^2 \kappa v_{\rm g}^2|\Phi_{\rm g}^0|^2 }\bigg[
K_{\rm l}\kappa^3 (U_{\rm g}\kappa+v_{\rm g}\sqrt{g^2-\kappa^2})
\nonumber\\
&+&g^2
\big[v_{\rm g}\kappa
[K_{\rm g}\sqrt{g^2-\kappa^2}+2(K_{\rm g}U_{\rm g}+D_{\rm g}v_{\rm g}|\Phi_{\rm g}^0|^2)]
-K_{\rm l}
[U_{\rm g}\kappa^2+2v_{\rm g}\sqrt{g^2-\kappa^2}
(\kappa-2v_{\rm g}|\Phi_{\rm g}^0|^2)]
\big]
\bigg].
\end{eqnarray}
Especially, at the CEP, 
\begin{eqnarray}
v^2&=&\frac{\kappa(D_{\rm g}v_{\rm g}+K_{\rm g}U_{\rm g}-K_{\rm l}U_{\rm g})}{v_{\rm g}}, \\
D&=& \frac{1}{2}\bigg[D_{\rm g}+\frac{K_{\rm g}U_{\rm g}}{v_{\rm g}}\bigg]>0.
\label{v2_supp}
\end{eqnarray}
This shows that the stability condition $v^2>0$ is satisfied when the nonlinear saturation $v_{\rm g}$ and the diffusion constant of the gain component $D_{\rm g}$ are large enough.

For completeness, we also provide the explicit form of the nonlinear couplings $t_\parallel=\sqrt{2}(t_{\rm l}-t_{\rm g})$ and $\lambda^\parallel_{\perp\perp}=[(\lambda^{\rm l}_{\rm ll}+\lambda^{\rm l}_{\rm lg}+\lambda^{\rm l}_{\rm gg})
-(\lambda^{\rm g}_{\rm ll}+\lambda^{\rm g}_{\rm lg}+\lambda^{\rm g}_{\rm gg})]/(2\sqrt{2})$ at the CEP:
\begin{eqnarray}
t_\parallel&=&\frac{\sqrt{2}U_{\rm g}\kappa}{v_{\rm g}},\\
\lambda^\parallel_{\perp\perp}
&=& - \frac{D_{\rm g}U_{\rm g}+(K_{\rm l}-K_{\rm g})v_{\rm g}}{2\sqrt{2}v_{\rm g}}.
\end{eqnarray}
Note how the sign of the KPZ-like coupling $\lambda^\parallel_{\perp\perp}$ depends on the details of the parameters in the original GP equation. This makes it possible for the effective coupling to take either sign, 
\begin{eqnarray}
\Gamma=\frac
{16\sqrt{2}
(K_{\rm l}-K_{\rm g}) U_{\rm g}v_{\rm g}
\big[
D_{\rm g}U_{\rm g}+(K_{\rm l}-K_{\rm g})v_{\rm g}
\big]
(U_{\rm g}^2+v_{\rm g}^2)\kappa^3
}
{(K_{\rm g}U_{\rm g}+D_{\rm g}v_{\rm g})^5}
\end{eqnarray}
leading to a distinct phases of matter, as we have discussed in the main text. 

\section{Dynamic renormalization group}
\label{appendixRG}

We extend the perturbative dynamic renormalization group method \cite{Hohenberg1977} (See also Ref. \cite{Medina1989} for the case on the KPZ equation.) to our coupled KPZ-like equation of motion (1). 
This is performed conveniently in the in-phase and out-of-phase basis,
\begin{eqnarray}
\delta\tilde \theta_{s}(\bm k,\omega) = \sum_{\alpha={\rm g,l}}{\mathcal U}_{s\alpha}\delta\theta_\alpha(\bm k,\omega), 
\tilde \xi_{s}(\bm k,\omega) = \sum_{\alpha={\rm g,l}}
{\mathcal U}_{s\alpha} \xi_\alpha(\bm k,\omega).
\end{eqnarray}
with 
\begin{eqnarray}
{\mathcal U}^\dagger\equiv
U^{\dagger}(\bm k=0)=(\bm u_\perp(\bm k=0),\bm u_\parallel(\bm k=0))^{\mathsf T}
=\frac{1}{\sqrt{2}}
\left(\begin{array}{cc}
1 & -1 \\
1 &  1
\end{array}\right),
\end{eqnarray}
transforming the kernel as,
\begin{eqnarray}
\tilde W(\bm k)
= {\mathcal U}W(\bm k){\mathcal U}^{\dagger}
=\left(\begin{array}{cc}
-D_{\perp\perp} \bm k^2 & \zeta - D_{\perp\parallel}\bm k^2 \\
-\frac{1}{\zeta}v^2 \bm k^2 &-\gamma - D_{\parallel\parallel} \bm k^2 
\end{array}\right),
\end{eqnarray}
with $\zeta=s_{\rm l}+s_{\rm g}, 
D_{\perp\perp}=(\nu_{\rm ll}+\nu_{\rm gg}+\nu_{\rm lg}+\nu_{\rm gl})/2, 
D_{\parallel\parallel} = (\nu_{\rm ll}+\nu_{\rm gg}-\nu_{\rm lg}-\nu_{\rm gl}) /2$, 
and $D_{\perp\parallel}=(-\nu_{\rm ll}+\nu_{\rm gg}+\nu_{\rm lg}-\nu_{\rm gl})/2$.
The equation of motion (1) is expressed in this basis as
($s=\perp,\parallel$),
\begin{eqnarray}
-i\omega\delta\tilde\theta_s(\bm k,\omega)
&=&\sum_{s'}\tilde W_{ss'}(\bm k)\delta\tilde\theta_{s'}(\bm k,\omega)
+t_s
\sum_{\bm k'}\int_{-\infty}^\infty\frac{d\omega'}{2\pi}
\delta\tilde\theta_{\parallel}(\bm k',\omega')
\delta\tilde\theta_{\parallel}(\bm k-\bm k',\omega-\omega')
\nonumber\\
&-&\sum_{s',s''}\lambda_{s's''}^s
\sum_{\bm k'}\int_{-\infty}^\infty\frac{d\omega'}{2\pi} [\bm k'\cdot(\bm k'-\bm k)]
\delta\tilde\theta_{s'}(\bm k',\omega')
\delta\tilde\theta_{s''}(\bm k-\bm k',\omega-\omega')
+\tilde\xi_s(\bm k,\omega).
\label{coupledKPZtri_supp}
\end{eqnarray}
We briefly note that the in-and-out-of phase basis used here is slightly different from the triangular basis introduced in the main text, where the present basis $\mathcal U^\dagger$ is momentum-independent. The former is more convenient since the massive nonlinear term only involves the out-of-phase fluctuations $\delta\tilde\theta_\parallel(\bm k,\omega)$, but we emphasize that the two representations are identical in the uniform limit $\bm k\rightarrow 0$. 


A simple power-counting analysis tells us that the diffusion constants $D_{\perp\perp},D_{\perp\parallel}$, and $D_{\parallel\parallel}$ are irrelevant when the velocity $v$ is fixed because the former are coefficients of higher spatial derivatives. However, the average of $D_{\perp\perp}$ and $D_{\parallel\parallel}$, i.e. $D=(D_{\perp\perp}+D_{\parallel\parallel})/2=(\nu_{\rm ll}+\nu_{\rm gg})/2$, turns out to be \textit{dangerously} irrelevant, in the sense that $D$ still affects the critical properties, as we have discussed in the main text. From here on, we ignore the parts that do not affect the critical properties, by putting $D_{\perp\perp}=D_{\parallel\parallel} = D$ and $D_{\perp\parallel}=0$.

Equation (\ref{coupledKPZtri_supp}) can be rewritten in terms of Green's function $\tilde G^0$ as,
\begin{eqnarray}
\delta\tilde\theta_s(\bm k,\omega)
&=&\sum_{s'}\tilde G^0_{ss'}(\bm k,\omega)\tilde\xi_{s'}(\bm k,\omega)
+\sum_{s'}\tilde G^0_{ss'}(\bm k,\omega) t_{s'} 
\sum_{\bm k'}\int_{-\infty}^\infty\frac{d\omega'}{2\pi}
\delta\tilde\theta_{\parallel}(\bm k',\omega')
\delta\tilde\theta_{\parallel}(\bm k-\bm k',\omega-\omega')
\nonumber\\
&-&\sum_{s',s'',s'''}\tilde G^0_{ss'}(\bm k,\omega)
\lambda_{s''s'''}^{s'}\sum_{\bm k'}\int_{-\infty}^\infty\frac{d\omega'}{2\pi} [\bm k'\cdot(\bm k'-\bm k)]
 \delta\tilde\theta_{s''}(\bm k',\omega')
\delta\tilde\theta_{s'''}(\bm k-\bm k',\omega-\omega'),
\end{eqnarray}
where the non-perturbative Green's function $\tilde G_{ss'}^0(\bm k,\omega)=([-i\omega\bm 1-\tilde W(\bm k,\omega)]^{-1})_{ss'}$ is given by,
\begin{eqnarray}
\tilde G^0(\bm k,\omega)=
\frac{1}{(\omega-\omega_-(\bm k))(\omega-\omega_+(\bm k))}
\left(\begin{array}{cc}
i\omega-\gamma - D\bm k^2 & -2\kappa \\
v^2 \bm k^2/(2\kappa) & i\omega-D \bm k^2
\end{array}\right).
\end{eqnarray}
Here, we have used the relation $\zeta=2\kappa$ that holds at the CEP. 
By iteratively substituting $\delta\tilde\theta_s(\bm k,\omega)$ into the right-hand side and taking the noise average, the perturbative corrections to the linearized theory can be conveniently computed by diagrammatic techniques. The obtained corrections are formalized into a dynamic renormalization group by integrating out only the fast modes with $\Lambda_c e^{-l}<|\bm k|<\Lambda_c$ and then rescaling the valuables according to 
\begin{eqnarray}
\bm r\rightarrow e^{l} \bm r, t\rightarrow e^{zl} t,\delta\tilde\theta_s\rightarrow e^{\chi l}\delta\tilde \theta_s.
\label{transform_supp}
\end{eqnarray}

\begin{figure}
\begin{center}
\includegraphics[width=0.85\linewidth,keepaspectratio]{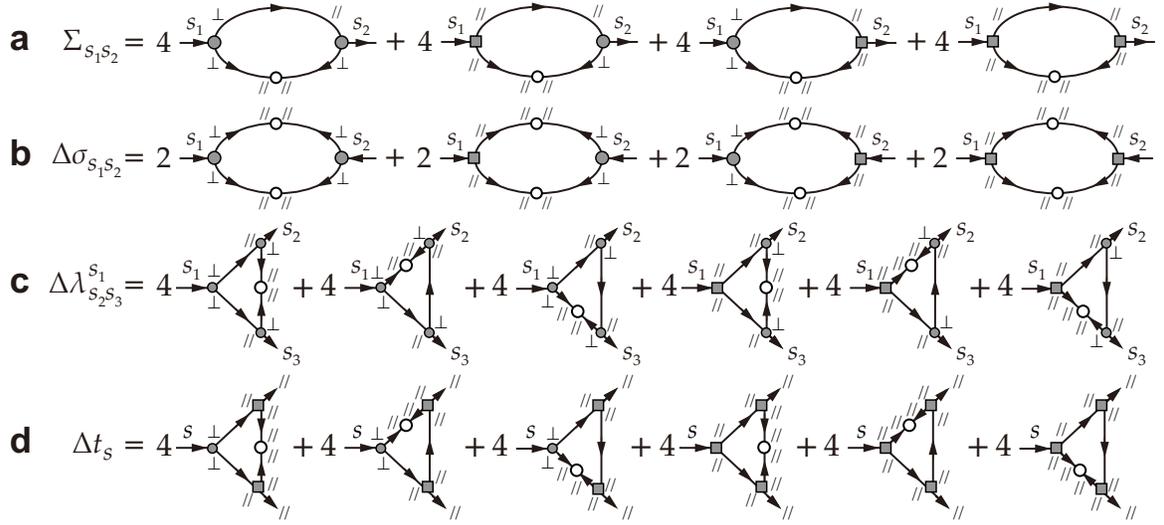}
\end{center}
\caption{
(Color online) 
Diagrammatic representation of nonlinear corrections.
(a) The self energy $\Sigma_{s_1s_2}$. 
(b) The noise kernel correction $\Delta\sigma_{s_1s_2}$.  
(c) The three-point vertex correction to the KPZ-like coupling $\Delta\lambda^{s_1}_{s_2s_3}$ and 
(d) the massive out-of-phase coupling $\Delta t_{s}$.
Here, the solid line is the Green's function $\tilde G_{ss'}^0$, the open circle represents the longitudinal noise strength $\sigma_{\parallel\parallel}$, the solid circle is the KPZ-like nonlinear coupling $\lambda^{s}_{s's''}$, and the solid square represents the massive out-of-phase coupling $t_s$.
}
\label{fig_diagram}
\end{figure}

The self energy $\Sigma_{ss'}$, defined as the correction to the Green's function (where $\delta\tilde\theta_s(\bm k,\omega)\equiv \sum_{s'}\tilde G_{ss'}(\bm k,\omega)\tilde\xi_{s'}(\bm k,\omega)$) 
by $\tilde G^{-1}_{ss'}(\bm k,\omega)=([\tilde G^0(\bm k,\omega)]^{-1})_{ss'}-\Sigma_{ss'}(\bm k,\omega)$,
is given diagrammatically by Fig. \ref{fig_diagram}(a) within the one-loop order. 
Here, we have dropped the terms that are obviously less relevant than the ones retained, by using the following properties at the CEP: (1) The most relevant propagator is the non-Hermitian-induced off-diagonal component $\tilde G^0_{\perp\parallel}$, hence, for the KPZ-like nonlinear couplings $\lambda^s_{s's''}$, the diagrams associated with $\tilde G^0_{\perp\parallel}$'s are the most relevant. 
(2) For the massive out-of-phase coupling $t_s$, the two out-going lines should be labeled ``$\parallel$''. Thus, they may only be connected to propagators $\tilde G^0_{\parallel\parallel}$ and $\tilde G^0_{\parallel\perp}$. (3) Since $\tilde G^0_{\parallel\parallel}$ is more relevant than $\tilde G^0_{\parallel\perp}$ (because $\kappa\rightarrow \infty$ at the fixed point), for massive nonlinearity $t_s$, the diagrams associated with $\tilde G^0_{\parallel\parallel}$'s are the most relevant.

Computing the self-energy shown in Fig. \ref{fig_diagram}(a) at $\omega=0$ (which is all we need for our purpose), given by,
\begin{eqnarray}
&&
\Sigma_{s_1s_2}(\bm k,\omega=0)
=
4\sigma_{\parallel\parallel}
\sum_{\bm q}\int_{-\infty}^\infty \frac{d\omega'}{2\pi}
\nonumber\\
&&\times
\bigg[
(-\lambda^{s_1}_{\perp\perp})(-\lambda^{\parallel}_{\perp s_2})
\Big[
\Big(\bm q+\frac{\bm k}{2}\Big)\cdot\Big(-\bm q+\frac{\bm k}{2}\Big)
\Big]
\Big[
\Big(-\bm q-\frac{\bm k}{2}\Big)\cdot\bm k
\Big]
\tilde G^0_{\perp\parallel}(\bm q+\frac{\bm k}{2},\omega')
\tilde G^0_{\perp\parallel}(-\bm q+\frac{\bm k}{2},-\omega')
\tilde G^0_{\perp\parallel}(-\bm q-\frac{\bm k}{2},-\omega')
\nonumber\\
&&\ \ \ \ \ 
+t_{s_1}(-\lambda^{\parallel}_{\perp s_2})
\Big[
\Big(-\bm q-\frac{\bm k}{2}\Big)\cdot\bm k
\Big]
\tilde G^0_{\parallel\parallel}(\bm q+\frac{\bm k}{2},\omega')
\tilde G^0_{\parallel\parallel}(-\bm q+\frac{\bm k}{2},-\omega')
\tilde G^0_{\perp\parallel}(-\bm q-\frac{\bm k}{2},-\omega')
\nonumber\\
&&\ \ \ \ \ 
+(-\lambda^{s_1}_{\perp\perp})t_{s_2}
\Big[
\Big(\bm q+\frac{\bm k}{2}\Big)\cdot\Big(-\bm q+\frac{\bm k}{2}\Big)
\Big]
\tilde G^0_{\perp\parallel}(\bm q+\frac{\bm k}{2},\omega')
\tilde G^0_{\perp\parallel}(-\bm q+\frac{\bm k}{2},-\omega')
\tilde G^0_{\parallel\parallel}(-\bm q-\frac{\bm k}{2},-\omega')
\nonumber\\
&&\ \ \ \ \ 
+ t_{s_1} t_{s_2} 
\tilde G^0_{\parallel\parallel}(\bm q+\frac{\bm k}{2},\omega')
\tilde G^0_{\parallel\parallel}(-\bm q+\frac{\bm k}{2},-\omega')
\tilde G^0_{\parallel\parallel}(-\bm q-\frac{\bm k}{2},-\omega')
\bigg],
\end{eqnarray}
we obtain, 
\begin{eqnarray}
\Sigma_{\perp\perp} (\bm k,\omega=0)
&=&\bigg[\frac{C_6}{4d}  -\frac{3C_8}{8d} \tilde\gamma \bigg] \Xi_\perp
D\bm k^2, \\
\Sigma_{\perp\parallel} (\bm k,\omega=0)
&=&\bigg[
\frac{C_4}{4}-\frac{C_6}{4} \tilde\gamma 
\bigg]\Pi_\perp
 + O(\bm k^2), \\
\Sigma_{\parallel\perp} (\bm k,\omega=0)
&=& \bigg[
-\frac{C_6}{2d}  
+\frac{3C_8}{4d}\tilde\gamma 
\bigg]\Xi_\parallel
\frac{v^2}{2\kappa}\bm k^2, \\
\Sigma_{\parallel\parallel} (\bm k,\omega=0)
&=&\bigg[
\frac{C_4}{4}-\frac{C_6}{4} \tilde\gamma 
\bigg]\Pi_\parallel 
-\bigg[
\frac{C_8}{16d}
+\frac{C_{10}}{8d}\tilde\gamma \bigg]\Gamma D\bm k^2, 
\end{eqnarray}
where 
\begin{eqnarray}
\Gamma
&=&\frac{t_{\parallel}\sigma_{\parallel\parallel}} {D^5}
(t_{\parallel} v^2 +4\kappa^2\lambda^\parallel_{\perp\perp}), \\
\Xi_\parallel&=&\frac{\kappa^2\lambda^\parallel_{\perp\perp}}{D^3v^4}
(t_{\parallel} v^2 +4\kappa^2\lambda^\parallel_{\perp\perp}), \\
\Xi_\perp &=&\frac{\kappa^2\lambda^\parallel_{\perp\perp}}{D^4v^2}
(t_{\perp} v^2 +4\kappa^2\lambda^\perp_{\perp\perp}), \\
\Pi_\parallel
&=&\frac{t_{\parallel}\sigma_{\parallel\parallel}} {D^3v^2}
(t_{\parallel} v^2 +4\kappa^2\lambda^\parallel_{\perp\perp}), \\
\Pi_\perp
&=&\frac{t_{\parallel}\sigma_{\parallel\parallel}} {D^3v^2}
(t_{\perp} v^2 +4\kappa^2\lambda^\parallel_{\perp\perp}).
\end{eqnarray}
Here, we have restricted ourselves to the vicinity of the CEP $\gamma=0$ where we expanded the Green's function in terms of $\gamma$, 
and have retained the lowest order correction in terms of $D$, since it is a (dangerous) irrelevant parameter that flows to zero as $l\rightarrow \infty$. 
Since the dressed Green's function $G$ is given by,
\begin{eqnarray}
G^{-1}(\bm k,\omega)
&=&\left(\begin{array}{cc}
-i\omega+D \bm k^2-\Sigma_{\perp\perp}(\bm k) 
& -2\kappa-\Sigma_{\perp\parallel}(\bm k) \\
\frac{1}{2\kappa}v^2 \bm k^2-\Sigma_{\parallel\perp}(\bm k) &-i\omega+\gamma + D \bm k^2 -\Sigma_{\parallel\parallel}(\bm k)
\end{array}\right)
\nonumber\\
&\equiv&\left(\begin{array}{cc}
-i\omega+(D+\Delta D_{\perp\perp}) \bm k^2 
& -2(\kappa+\Delta \kappa) \\
\frac{1}{2\kappa}(v^2+\Delta v^2) \bm k^2 &-i\omega+(\gamma +\Delta \gamma)+ (D+\Delta D_{\parallel\parallel}) \bm k^2 
\end{array}\right),
\end{eqnarray}
these self energy corrections give the nonlinear correction $(\Delta v,\Delta\kappa,\Delta D,\Delta\gamma)$ to the valuables $(v,\kappa,D,\gamma)$. The correction to the diffusion constant is given by $\Delta D=(\Delta D_\perp+\Delta D_\parallel)/2$.


Now, let us examine which effective couplings are the most relevant. Since the transformation (\ref{transform_supp}) changes the parameters to 
\begin{eqnarray}
\kappa\rightarrow \kappa e^{zl},
v\rightarrow v e^{(z-1)l}, D\rightarrow D e^{(z-2)l}, 
\sigma_{\parallel\parallel}\rightarrow \sigma_{\parallel\parallel} e^{(z-d-2\chi)l},
\lambda^{s}_{s's''}\rightarrow \lambda^{s}_{s's''} e^{(z-2+\chi)l},
t_s\rightarrow t_s e^{(z+\chi)l},
\end{eqnarray}
the effective couplings change as, \end{widetext}
\begin{eqnarray}
\Gamma &\rightarrow& e^{(8-d)l}\Gamma, \\
(\Xi_\parallel,\Xi_\perp)& \rightarrow & e^{(6-d)l}  (\Xi_\parallel,\Xi_\perp),  \\
(\Pi_\parallel,\Pi_\perp) &\rightarrow& e^{(4-d)l} 
(\Pi_\parallel,\Pi_\perp).
\end{eqnarray}
This tells us that the effective coupling $\Gamma$ is more relevant than all other effective couplings, with upper critical dimension $d_c=8$.

We note that the most relevant coupling $\Gamma$ in $\Sigma_{\parallel\parallel}$ originates from the third and fourth diagrams in Fig. \ref{fig_diagram}(a). The third diagram involves both $\tilde G^0_{\perp\parallel}$ and $\tilde G^0_{\parallel\parallel}$, while 
the last only includes $\tilde G^0_{\parallel\parallel}$,
 which means that for the latter nonlinear correction is solely determined by the dynamics of the out-of-phase motion. 
However, note that $\tilde G^0_{\parallel\parallel}$ in the triangular basis  introduced in the main text ($\bm u_\perp(\bm k),\bm u_\parallel(\bm k)$) (i.e. the transverse and longitudinal basis) is given by, 
\begin{eqnarray}
\tilde G^0_{\parallel\parallel}(\bm k,\omega)
=\bar G^0_{\parallel\parallel}(\bm k,\omega)
+\frac{iv|\bm k|}{2\kappa}\bar G^0_{\perp\parallel}(\bm k,\omega).
\end{eqnarray}
The second term converts the longitudinal fluctuations to the transverse (Goldstone) mode, and has the strongest singularity. This contributes to the resulting singular form of effective coupling $\Gamma\propto D^{-5}$, which
is the origin of the anomalously high upper critical dimension.



Below, we restrict ourselves to spatial dimensions close to the upper critical dimension $d_c=8$, where $\Gamma$ is the only relevant term.
In such case, the self energy reduces to,
\begin{eqnarray}
\Sigma_{\parallel\parallel}(\bm k,\omega=0)
&=&
-\bigg[
\frac{C_8}{16d}
+\frac{C_{10}}{8d}\tilde\gamma \bigg]\Gamma D\bm k^2,
\end{eqnarray} 
and $\Sigma_{\perp\perp}=\Sigma_{\perp\parallel}=\Sigma_{\parallel\perp}=0$, which gives
\begin{eqnarray}
\Delta D=\bigg[
\frac{C_8}{32d}
+\frac{C_{10}}{16d}\tilde\gamma \bigg]\Gamma,
\end{eqnarray}
and $\Delta \gamma=\Delta\kappa=\Delta v=0$.
The flow equations are thus given by,
\begin{eqnarray}
\frac{d\gamma}{dl}
&=&z\gamma, 
\label{betagamma_supp}
\\
\frac{d\kappa}{dl}
&=&z\kappa, 
\label{betakappa_supp}
\\
\frac{dv}{dl}
&=&
(z-1)v, 
\label{betav_supp}
\\
\frac{d D}{dl}
&=&\Big[z-2+\ \Big(
\frac{C_8}{32d} 
+\frac{C_{10}}{16d}
\tilde\gamma 
\Big)\Gamma \Big]D.
\label{betaD_supp}
\end{eqnarray}

The noise kernel correction $\Delta\sigma_{ss'}$ and the three-point vertex corrections $\Delta\lambda^s_{s's''},\Delta t_s$ can be analyzed similarly by computing the diagrams shown in Figs. \ref{fig_diagram}(b), (c), and (d), respectively. 
It turns out that these diagrams only give nonlinear effective couplings that are less relevant than $\Gamma$, with the maximum upper critical dimension being $d_c=6$. Ignoring such less-relevant terms, the flow equations for the noise kernel and the three-point vertices are given by,
\begin{eqnarray}
\frac{d\sigma_{ss'}}{dl}
&=&
(z-d-2\chi )\sigma_{ss'},\\
\frac{d\lambda^{s_1}_{s_2s_3}}{dl}
&=& 
(z-2 +\chi )\lambda^{s_1}_{s_2s_3},\\
\frac{d t_s}{dl}
&=& 
(z+\chi ) t_s.
\label{betats_supp}
\end{eqnarray}
Combining the flow equations for $\gamma,\kappa,v,D,\sigma_{\parallel\parallel},\lambda^\parallel_{\perp\perp}$, and $t_\parallel$ (Eqs. (\ref{betagamma_supp})-(\ref{betats_supp})), we obtain the flow equations for $\tilde \gamma=\gamma/D$ (Eq. (\ref{betakappa})), $A=\kappa^2\sigma_{\parallel\parallel}/(Dv^2)$ (Eq. (\ref{betaA})), and the effective coupling $\Gamma$ (Eq. (\ref{betaGamma})) presented in the main text.

\end{appendix}





\end{document}